\newcommand*{\ATLASLATEXPATH}{./}
\newcommand*{\FIGUREPATH}{./}
\documentclass[UKenglish,texlive=2011,txfonts,texmf,cernpreprint,maketitle=false]{\ATLASLATEXPATH atlasdoc}
\pdfoutput=1
\usepackage{\ATLASLATEXPATH atlaspackage}
\usepackage[articletitle=true]{\ATLASLATEXPATH atlasbiblatex}
\usepackage{\ATLASLATEXPATH atlasphysics}

\graphicspath{{logos/}{figs/}}

\hypersetup{pdftitle={ATLAS draft},pdfauthor={The ATLAS Collaboration}}
\AtlasTitle{\boldmath Modelling ${Z\to\tau\tau}$ processes in ATLAS \linebreak with $\mathbf{\tau}$-embedded 
${Z\to\mu\mu}$ data\unboldmath }

\author{The ATLAS Collaboration}

\PreprintIdNumber{CERN-PH-EP-2015-130}

\AtlasJournal{JINST}

\AtlasAbstract{This paper describes the concept, technical realisation and validation of a largely data-driven method
to model events with $Z\to\tau\tau$ decays. In $Z\to\mu\mu$ events selected from proton--proton collision data
recorded at $\sqrt{s}=8$~TeV with the ATLAS experiment at the LHC in 2012, the $Z$ decay muons are replaced by
$\tau$ leptons from simulated $Z\rightarrow\tau\tau$ decays at the level of reconstructed tracks and calorimeter cells.
The $\tau$ lepton kinematics are derived from the kinematics of the original muons. Thus, only the well-understood
decays of the $Z$ boson and $\tau$ leptons as well as the detector response to the $\tau$ decay products are obtained
from simulation. All other aspects of the event, such as the $Z$ boson and jet kinematics as well as effects from multiple
interactions, are given by the actual data.
This so-called {\it $\tau$-embedding} method is particularly relevant for Higgs boson searches and analyses in $\tau\tau$
final states, where $Z\rightarrow\tau\tau$ decays constitute a large irreducible background that cannot be obtained directly
from data control samples.
In this paper, the relevant concepts are discussed based on the implementation used in the ATLAS Standard Model $H\to\tau\tau$ analysis of the full datataset recorded during 2011 and 2012.
}

\begin{document}

\maketitle
\section{Introduction}
\label{sec:intro}
 %
 %
The experimental sensitivity of searches for (and eventually studies of) Higgs bosons in $\tau\tau$ final states
at the LHC is driven by analyses of intricate event signatures that are not restricted to the Higgs candidate decay products.
For example, the missing transverse momentum enters the reconstruction of the di-$\tau$ invariant mass $m_{\tau\tau}$,
which is a key quantity in these analyses. The shape of the reconstructed $m_{\tau\tau}$ distribution also depends on
the boost of the $\tau\tau$ system and thus on the presence and kinematics of additional jets in the event.
In addition, details of the final-state topology are used to define event categories, for example based on vector-boson
fusion topologies characterised by two high-energy jets with large rapidity separation, and recent ATLAS
analyses~\cite{Aad:2015vsa} also combine them into multivariate classifiers to extract the Higgs boson signal.

In these analyses, events with $Z/\gamma^*\rightarrow\tau\tau$ decays constitute a large irreducible background,
and thus a reliable and detailed model of these processes is a critical ingredient. In view of the complexity of the
relevant event properties it is highly desirable to rely as little as possible on simulation; moreover, it has been shown
in dedicated measurements~\cite{Aad:2013ysa,Aad:2014xaa,Aad:2014dta} that existing Monte Carlo simulations of
$Z$+jets events need to be corrected in order to model the data.
Ideally the model would be obtained directly from the collision data.
However, due to background contributions, e.g. from events with other objects misidentified as $\tau$ decays,
it is difficult to select a sufficiently pure $Z/\gamma^*\rightarrow\tau\tau$ sample from the data, and
doing so without also including Higgs boson decays to $\tau$ lepton pairs is conceptually impossible.

$Z/\gamma^*\rightarrow\tau\tau$ events can still be modelled in a largely data-driven way by using
$Z/\gamma^*\rightarrow\mu\mu$ events as a starting point.\footnote{For simplicity, these processes
are hereafter denoted by $Z\to\tau\tau$ and $Z\to\mu\mu$, respectively.
}
Except for effects due to the difference in muon and $\tau$ lepton masses, the two processes are kinematically
identical assuming lepton universality. In particular the kinematics of the $Z$ boson and additional jets in the event
are independent of the $Z$ decay mode. By requiring two isolated, high-energy muons with opposite charge,
$Z\rightarrow\mu\mu$ decays can be selected from the data with high efficiency and purity,
and due to the small muon mass and correspondingly small Higgs--muon coupling, the $H\rightarrow\mu\mu$
contamination is expected to be negligible for all practical purposes. The detector response
to the $Z$ decay muons can be removed from the data events and replaced by corresponding information
for $\tau$ leptons from simulated $Z\rightarrow\tau\tau$ decays, where the $\tau$ kinematics are derived from the kinematics
of the original muons (taking into account both the $\tau$--$\mu$ mass difference and the $\tau$--$\tau$ spin correlation).
This substitution results in a $Z\rightarrow\tau\tau$ event model where only the well-understood decays of the $Z$ boson
and $\tau$ leptons and the detector response to the $\tau$ lepton decay products are obtained from the simulation. All other
aspects of the event -- including, for example, the kinematics of the $Z$ boson and additional jets, the underlying event as
well as effects from multiple interactions -- are directly taken from the data. The simulated and collision-data information are
combined based on reconstructed tracks and calorimeter cells, followed by a re-reconstruction of the resulting hybrid events.
In the following, this technique is referred to as {\it embedding} of simulated $Z\rightarrow\tau\tau$ decays in
$Z\rightarrow\mu\mu$ data events (or, in short, {\it $\tau$ embedding}). It has been used in all $H\to\tau\tau$
searches by ATLAS~\cite{Aad:2011rv,Aad:2012mea,Aad:2012cfr,Aad:2014vgg} to date, including the most recent
analysis~\cite{Aad:2015vsa} establishing evidence for this decay. Corresponding CMS
analyses~\cite{Chatrchyan:2014nva,Chatrchyan:2012vp} have applied a similar technique.
In addition, the method was adapted to single-$\tau$ processes for use in the analysis of
$W\rightarrow\tau\nu_\tau$ decays~\cite{Aad:2011fu} and searches for charged Higgs
bosons~\cite{Aad:2012tj,Aad:2014kga}.

This paper describes the concept, technical realisation and validation of the $\tau$ embedding corresponding to the
implementation used in the ATLAS $H\to\tau\tau$ analysis~\cite{Aad:2015vsa} of the full $pp$ collision dataset recorded
during 2011 and 2012 at centre-of-mass energies of $\sqrt{s}=7$~TeV and $\sqrt{s}=8$~TeV, respectively.
The method is valid for all $\tau$ lepton decay channels. However, here the discussion and examples focus on final states
where one of the $\tau$ leptons decays leptonically and the other one hadronically, also referred to below as
the {\it lepton-hadron} $\tau\tau$ decay mode. This corresponds to the most sensitive $H\to\tau\tau$ channel and
tests the embedding of both the leptonic and hadronic $\tau$ decays.
After a description of the ATLAS detector and the final-state reconstruction algorithms in Section~\ref{sec:exp},
Section~\ref{sec:sampsel} provides an overview of the relevant event samples and selections.
Section~\ref{sec:method} outlines the concept and implementation of the $\tau$-embedding
method. Studies to validate the procedure and associated systematic uncertainties are discussed in
Section~\ref{sec:validation}. A summary and conclusions are given in Section~\ref{sec:conclusions}.
%
\section{Experimental setup}
\label{sec:exp}
%
%
\subsection{The ATLAS detector}
\label{subsec:atlas}
%
The ATLAS detector~\cite{Aad:2008zzm} at the LHC covers nearly the entire solid angle around the collision
point. It consists of an inner tracking detector surrounded by a thin superconducting solenoid, electromagnetic
and hadronic calorimeters, and a muon spectrometer incorporating three large superconducting toroid magnets,
each with eight coils.
The inner-detector
system (ID) is immersed in a
2~T axial magnetic field and provides charged-particle tracking in the pseudorapidity range\footnote{ATLAS
uses a right-handed coordinate system with its origin at the nominal interaction point (IP)
in the centre of the detector and the $z$-axis along the beam pipe.
The $x$-axis points from the IP to the centre of the LHC ring,
and the $y$-axis points upwards.
Cylindrical coordinates $(r,\phi)$ are used in the transverse plane,
$\phi$ being the azimuthal angle around the beam pipe.
The pseudorapidity is defined in terms of the polar angle $\theta$ as $\eta = -\ln \tan(\theta/2)$.
Angular distance is measured in units of $\Delta R \equiv \sqrt{(\Delta\eta)^{2} + (\Delta\phi)^{2}}$.
}
$|\eta| < 2.5$.
The high-granularity silicon pixel detector covers the vertex region and typically provides three measurements per track.
It is followed by the silicon microstrip tracker which usually provides four two-dimensional measurement points per track.
These silicon detectors are complemented by the transition radiation tracker, which enables radially extended track
reconstruction up to $|\eta| = 2.0$. The transition radiation tracker also provides electron identification information based on
the fraction of hits (typically 30 in total) above a higher energy-deposit threshold corresponding to transition radiation. The
calorimeter system covers the pseudorapidity range $|\eta| < 4.9$. Within the region $|\eta|< 3.2$, electromagnetic
calorimetry is provided by barrel and end cap high-granularity lead/liquid-argon (LAr) electromagnetic calorimeters,
with an additional thin LAr presampler covering $|\eta| < 1.8$, to correct for energy loss in material
between the interaction vertex and the calorimeters. Hadronic calorimetry is provided by the steel/scintillating-tile
calorimeter, segmented into three barrel structures within $|\eta| < 1.7$, and two copper/LAr hadronic endcap calorimeters.
The solid angle coverage is completed with forward copper/LAr and tungsten/LAr calorimeter modules optimised for
electromagnetic and hadronic measurements respectively. The muon spectrometer (MS) comprises separate trigger and
high-precision tracking chambers measuring the deflection of muons in a magnetic field generated by superconducting
air-core toroids. The precision chamber system covers the region $|\eta| < 2.7$ with three layers of monitored drift tubes,
complemented by cathode strip chambers in the forward region, where the background is highest. The muon trigger system
covers the range $|\eta| < 2.4$ with resistive plate chambers in the barrel, and thin gap chambers in the endcap regions. A
three-level trigger system is used to select interesting events~\cite{TriggerPaper}. The Level-1 trigger is implemented in
hardware and uses a subset of detector information to  reduce the event rate to a design value of at most  75~kHz.
This is followed by two software-based trigger levels which together reduce the event rate to about 400~Hz.

\subsection{Final-state reconstruction}
\label{sec:objects}
%
Muon candidates are reconstructed using an
algorithm~\cite{Aad:2014rra}
that combines information from the ID and the MS.  The distance between the
$z$-position of the point of closest approach of the muon inner-detector track to
the beam-line and the $z$-coordinate of the primary vertex\footnote{The primary vertex is the proton--proton vertex candidate with the highest
sum of the squared transverse momenta of all associated tracks.}
is required to be less
than 1 cm. This requirement reduces the contamination due to cosmic-ray muons
and beam-induced backgrounds. Muon quality criteria such as inner-detector hit
requirements are applied in order to achieve a precise measurement of the muon
momentum and reduce the misidentification rate.  Muons are required to have a
momentum in the transverse plane $\pT >10\GeV$ and a pseudorapidity of $|\eta|<$~2.5.
Isolation requirements on close-by tracks and energy depositions in the calorimeter
are applied in order to distinguish prompt muons from other candidates originating e.g.\ from hadronic showers.

Electron candidates are reconstructed from energy clusters in the
electromagnetic calorimeters matched to a track in the ID. They are required to
have a transverse energy, $\ET = E\sin\theta$, greater than $15\GeV$, be within the
pseudorapidity range $|\eta|<2.47$ and satisfy the {\it medium} shower shape and
track selection criteria defined in Ref.~\cite{Aad:2014fxa}.  Candidates found
in the transition region between the end-cap and barrel calorimeters ($1.37<|\eta|<1.52$) are not considered.
Like for the muons, isolation criteria are applied to suppress non-prompt candidates originating e.g.\ from hadronic showers.

Jets are reconstructed using the anti-$k_t$ jet clustering
algorithm~\cite{Antikt1,Antikt2} with a radius parameter $R=0.4$, taking topological energy
clusters~\cite{ATL-LARG-PUB-2008-002} in the calorimeters as inputs. Jet
energies are corrected for the contribution of multiple interactions using a
technique based on jet area~\cite{Cacciari:2007fd} and are calibrated using $\pT$-
and $\eta$-dependent correction factors determined from simulation and
data~\cite{Aad:2011he,Aad:2014bia,Aad:2012vm}.
Jets are required to be reconstructed
in the range $|\eta|<4.5$ and to have $\pT > 30\GeV$.
To reduce the contamination by jets from additional $pp$ interactions in the same or
neighbouring bunch crossings (pile-up), tracks originating from the primary vertex
must contribute a large fraction of the $\pT$ when summing the scalar $\pT$ of all tracks associated with the
jet.  This jet vertex fraction (JVF) is required to be at least 50\% for jets
with $\pT < 50\GeV$ and $| \eta | < 2.4$. Jets with no associated tracks
are retained.

Hadronically decaying $\tau$ leptons are reconstructed starting from clusters of
energy depositions in the electromagnetic and hadronic calorimeters. The
$\tau_{\mathrm{had}}$\footnote{In the following, the $\tau_{\mathrm{had}}$ symbol always refers to the visible
decay products of the hadronic $\tau$ decay.} reconstruction is seeded by the anti-$k_t$
jet-finding algorithm with a radius parameter $R = 0.4$. Tracks with $\pt >1\GeV$ within a cone of size
$\Delta R=0.2$ around the cluster barycentre are assigned to the
$\tau_{\mathrm{had}}$ candidate.
Its momentum is calculated from the topological energy clusters associated with the jet seed after applying a
dedicated $\tau_{\mathrm{had}}$  energy calibration. The $\tau_{\mathrm{had}}$ charge is determined from the
sum of the charges of the associated tracks.
The rejection of jets is provided in a separate identification step using
discriminating variables based on tracks with $\pt > 1\GeV$ and the energy deposited in calorimeter cells
found in the core region ($\Delta{}R < 0.2$) and in the region $0.2 <
\Delta{}R < 0.4$ around the $\tau_{\mathrm{had}}$ candidate's direction. Such discriminating
variables are combined in a boosted decision tree and three working points,
labelled {\em tight}, {\em medium} and {\em loose}~\cite{Aad:2014rga},
are defined, corresponding to different $\tau_{\mathrm{had}}$ identification efficiency
values. In the studies presented in this paper,
$\tau_{\mathrm{had}}$ candidates with $p_\mathrm{T}>20\GeV$ and
$|\eta|<2.47$ are used. The $\tau_{\mathrm{had}}$ candidates are required to have one or three reconstructed tracks
with a total charge of $\pm{}1$ and to satisfy the medium criteria, which provide an identification efficiency of the order of 55--60\%.  Dedicated
criteria~\cite{Aad:2014rga} to separate $\tau_{\mathrm{had}}$ candidates from
misidentified electrons are also applied, with a selection efficiency for true
hadronic $\tau$ decays of $95\%$.  The probability to misidentify a jet with
$p_\mathrm{T}>20\GeV$ as a $\tau_{\mathrm{had}}$ candidate is typically 1--2\%.

Following their reconstruction, candidate leptons, hadronically decaying $\tau$ leptons
and jets may point to the same energy deposits in the calorimeters. Two reconstructed objects
are considered to overlap if their separation $\Delta R$ is smaller than 0.2.
Such overlaps are removed by selecting objects in the following order of
priority (from highest to lowest): muons, electrons, $\tau_{\mathrm{had}}$, and jet candidates.
Objects with lower priority are discarded when overlapping with another object with higher priority.
The leptons that are considered in overlap removal with $\tau_{\mathrm{had}}$ candidates need
only to satisfy looser criteria than those defined above, to reduce
misidentified $\tau_{\mathrm{had}}$ candidates from leptons. The \pt\ threshold of muons
considered in overlap removal is also lowered to $4\GeV$.

The missing transverse momentum (with magnitude \MET)
is reconstructed using the energy deposits in calorimeter cells calibrated
according to the reconstructed physics objects ($e$, $\gamma$, $\tau_{\mathrm{had}}$, jets and
$\mu$) with which they are
associated~\cite{Aad:2012re}.
The transverse momenta of reconstructed muons are included in the \MET\
calculation, with the energy deposited by these muons in the calorimeters taken
into account.  The energy from calorimeter cells not associated with any physics
object is scaled  according to a soft-term vertex fraction and also included in the
\MET\ calculation. This fraction is the ratio of the summed scalar \pt\ of
tracks from the primary vertex not matched with objects to the summed scalar \pt\ of
all tracks in the event also not matched to objects. This method allows a better
reconstruction of the \MET\ in high pile-up
conditions~\cite{ATLAS-CONF-2013-082}.
%
\section{Data samples and event selection}
\label{sec:sampsel}
%
%
\subsection{Event samples}
\label{subsec:samples}
%
The studies presented in this paper are based on data recorded with ATLAS during the 2012 LHC run at a
proton--proton centre-of-mass energy \mbox{$\sqrt{s}=8\TeV$}. After data-quality requirements, these
correspond to an integrated luminosity of 20.3~fb$^{-1}$.

For the validation of the $\tau$-embedding procedure, samples of Monte Carlo simulated (MC) events with
$Z\to\mu\mu$ and $Z\to\tau\tau$ decays are used as input or as reference, respectively.
Simulated events are produced with the \textsc{Alpgen}~\cite{alpgen} event generator employing the \textsc{MLM}
matching scheme~\cite{MLM} between the hard process (calculated with leading-order matrix elements for up to five
partons) and the parton shower. The \textsc{Cteq6L1} parameterisation of the parton distribution
functions~\cite{Pumplin:2002vw} is used and the \textsc{Pythia8} program~\cite{Sjostrand:2007gs} provides the
modelling of the parton shower, the hadronisation and the underlying event.
 A full simulation of the ATLAS detector response~\cite{Aad:2010ah} using the \textsc{Geant4}
program~\cite{Agostinelli:2002hh} is performed. In addition,
events from minimum-bias interactions are simulated using the \textsc{AU2}~\cite{atlasmctunes} tuning
of \textsc{Pythia8}. They are overlaid with the simulated signal and background events according to the
luminosity profile of the recorded data. The contributions from these pile-up interactions are simulated both
within the same bunch crossing as the hard-scattering process and in neighbouring bunch crossings. Finally, the
resulting simulated events are processed through the same reconstruction programs as the data.

In the simulation of the $Z\to\tau\tau$ decays that are embedded into the $Z\to\mu\mu$ input events as described in
Section~\ref{subsec:proc}, the $\tau$ decay products are generated using \textsc{Tauola}~\cite{Was:2000st}, and
\textsc{Photos}~\cite{Davidson:2010ew} provides photon radiation from charged leptons.

\newpage
From these datasets, the following event samples are derived:

\begin{itemize}
\item Replacing the muons from recorded $Z\to\mu\mu$ data events with $\tau$ leptons from simulated $Z\to\tau\tau$
decays as described in Section~\ref{subsec:proc} results in {\it $\tau$-embedded} data, which are the standard event
samples used in physics analyses to model $Z\to\tau\tau$ processes.
\item  {\it $\mu$-embedded} data are obtained by using simulated $Z\to\mu\mu$ decays instead of $Z\to\tau\tau$ decays
to replace the muons in the $Z\to\mu\mu$ input data events. These make it possible to study systematic effects of the
embedding procedure in comparatively simple final states. While the $\tau$-embedded samples are based on the full 2012
dataset, the $\mu$-embedded validation is restricted to a subset corresponding to an integrated luminosity of 1.0~fb$^{-1}$.
\item  Using simulated instead of data $Z\to\mu\mu$ events as input yields $\mu$- or $\tau${\it -embedded MC} samples.
These can then be compared to direct simulations of these processes.
\item In order to study effects originating from the reconstruction of the input muons as well as of final-state radiation,
alternative embedded MC samples are produced, where the kinematics of the embedded objects are derived from the
generator-level muons instead of the reconstructed momenta. In the following, this is referred to as
{\it generator-seeded embedding}, as opposed to the standard {\it detector-seeded} procedure.
\end{itemize}

\subsection{Event selection}
\label{subsec:selection}
%
For the studies presented below, events are selected from one or several of the samples listed in
Section~\ref{subsec:samples} using one of the following sets of criteria. In all cases, standard quality criteria are applied
to ensure a fully operational detector and well-reconstructed events.

\begin{itemize}

\item
{\it $Z\to\mu\mu$ selection:}\\
Collision events are selected using a combined dimuon trigger, 	
with $\pt$ thresholds of 18~\GeV\ for the leading muon and 8~\GeV\
for the sub-leading muon, or a single-muon trigger ($\pt(\mu)>$ 24~\GeV). Only events with at least two good-quality
muons (cf.~Section~\ref{sec:objects}) are accepted.
The leading (sub-leading) muon is required to fulfil 
$\pt(\mu)>$ 20~(15)~\GeV.
Both muons must be isolated in the ID, which is ensured by requiring the scalar sum of other track transverse momenta
in an isolation cone of size $\Delta R=0.4$ to be smaller than 20\% of the muon transverse momentum
($I(\pt,0.4)/\pt(\mu)<0.2$). Only
events containing at least one such opposite-charge muon pair with an invariant mass \mbox{$m_{\mu\mu}>$ 40~\GeV} are
considered.

\item {\it $Z\to\tau\tau$ selection:}\\
The {$\tau\tau$ selection} is adopted from the $H\to\tau\tau$ lepton-hadron-channel analysis documented in
Ref.~\cite{Aad:2015vsa}.
Both in simulated and recorded data samples, single-electron or single-muon triggers with a lepton $\pt$ threshold
of 24~GeV are used to select events, in which exactly one $\tau$ candidate with $\pt(\tau_{\mathrm{had}})>$ 20~GeV
fulfilling the medium identification criteria and either exactly one electron or exactly one muon
with $\pt(e/\mu)>$ 26~GeV are required. In addition to a track isolation of $I(\pt,0.4)/\pt(e/\mu)<0.06$, a calorimeter
isolation of $I(E_T,0.2)/\pt(e/\mu)<0.06$ is applied to the leptons,  i.e.~the scalar sum of the transverse energy deposited
in calorimeter cells within $\Delta R<0.2$ not associated with the candidate is calculated and required to be smaller
than 6\% of the total transverse momentum of the muon or the total transverse energy of the electron.

\item {\it Boosted $Z$-enriched selection:}\\
The $H\to\tau\tau$ lepton-hadron-channel analysis documented in Ref.~\cite{Aad:2015vsa} considers two signal event categories: a {\it VBF} category enriched in vector-boson fusion Higgs production events and a {\it boosted} category targeting mainly events with high-$\pt$ Higgs bosons produced via gluon-gluon fusion. For the boosted category, a corresponding $Z$-enriched control sample is defined, which is adopted here to illustrate the $\tau$-embedding performance within physics analyses, see Section~\ref{subsec:physval}. 
This sample includes events that pass the $Z\to\tau\tau$ selection described above but fail the VBF category selection detailed in Ref.~\cite{Aad:2015vsa}. 
In addition, the $\pt$ of the $Z$ candidate reconstructed from the vector sum of momenta of the visible $\tau$ decay products and the
missing transverse momentum is required to exceed 100 GeV.
In order to further enhance the fraction of $Z$ events, 
$W$ decays are suppressed by considering only events with a transverse
mass\footnote{$m_T=\sqrt{2\cdot \pt(\ell)\cdot \MET\cdot\left(1-\cos\Delta\phi\right)}$; $\Delta\phi$
is the azimuthal angle between the directions of the electron or muon $\ell$ and the missing transverse momentum vector.
}
$m_T<$ 40~\GeV. Potential contamination by Higgs signal events is avoided by requiring the invariant mass
$m_{\tau\tau}^{\mathrm{MMC}}$ of the $\tau\tau$ pair not to exceed 110~GeV.
This mass is reconstructed from the visible $\tau$ decay products and the missing transverse momentum with the so-called
missing mass calculator (MMC)~\cite{Elagin:2010aw}. 
\end{itemize}
%
\section{Embedding}
\label{sec:method}
%
In the following, the $\tau$ embedding method is described in more detail. Special properties of the resulting event samples
and embedding-specific systematic uncertainties are also discussed.

\subsection{Procedure}
\label{subsec:proc}
\begin{figure}[h!]
    \centering
    \includegraphics[width=0.65\textwidth]{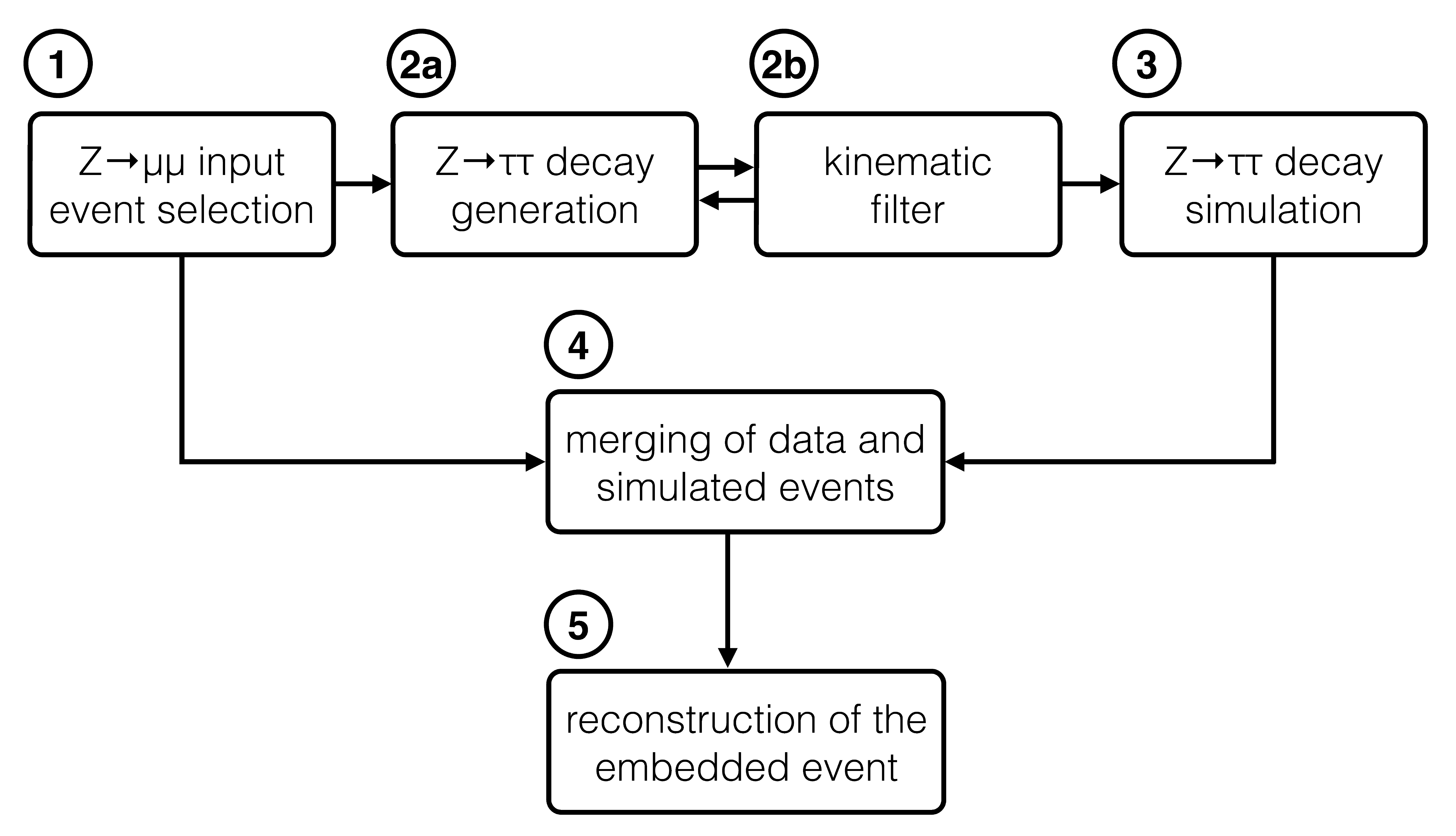}
    \caption{
             Flowchart of the embedding procedure.
        }
    \label{fig:EmbFlow}
    \vspace*{-0.3cm}
\end{figure}

%
The $\tau$ embedding procedure can be separated into five consecutive steps as depicted in the flowchart shown in
Figure~\ref{fig:EmbFlow}. After selecting the $Z\to\mu\mu$ input event, a corresponding  $Z\rightarrow\tau\tau$ decay
is generated and passed to a full detector simulation. The muons in the input event are then replaced by the $\tau$ leptons
from the simulated $Z$ decay. As a final step, a re-reconstruction of the resulting hybrid event is necessary, since it would
be insufficient  to combine the event information at the level of fully reconstructed physics objects. For example, the additional
calorimeter energy depositions from pile-up events can change the results of the \MET\ reconstruction, and the identification of
hadronic $\tau$ decays is particularly sensitive to the details of the calorimeter response. In contrast, corresponding effects on
the reconstruction of charged-particle tracks from the individual tracking detector hits are expected to be negligible for the
data-taking conditions and the phase space relevant to the Higgs analyses of the 8 TeV data. Therefore, the embedding
procedure is performed at the level of calorimeter cells and reconstructed tracks, as described in more detail below.\\
\begin{enumerate}
\item{Selection of the $Z\rightarrow\mu\mu$ input events from the collision data:}\\
Input events for the embedding procedure are obtained according to the $Z\to\mu\mu$  selection described in
Section~\ref{subsec:selection}. For events with more than two muon candidates, all possible
oppositely-charged pairs with a common vertex
are formed, and the muon pair with $m_{\mu\mu}$ closest to the $Z$ boson mass is
chosen as the $Z\to\mu\mu$ candidate decay products.\\ 
\item{Generation of a corresponding $Z\to\tau\tau$ decay:}
\begin{enumerate}
\item{Substitution of muons with $\tau$ leptons and subsequent $\tau$ decays:}\\
From the selected muons in a collision data event, the four-momenta of a corresponding $Z\rightarrow\tau\tau$ decay
are derived: the production vertex of the $\tau$ leptons is set to the common production vertex of the reconstructed
muon pair, and each muon is then replaced by a $\tau$ lepton. The $\tau$ four-momenta are rescaled
according to $$p_\tau=\sqrt{E_\mu^{2}-m_\tau^2},$$ thus keeping the energy $E_\mu$ unchanged but replacing
the muon mass with the $\tau$ mass $m_\tau$.

The resulting $Z\rightarrow\tau\tau$ kinematics as obtained from the $Z\rightarrow\mu\mu$ events is processed with
{\sc Tauola} and {\sc Photos}. Here, the decay of each $\tau$ lepton pair by {\sc Tauola} takes into account the polarisation
and spin correlations of the $\tau$ leptons. The $Z$ polarisation, however, depends on the parton configuration of the
initial state, which is not directly available here. During the generation of the decays, {\sc Tauola} therefore assumes an
average polarisation of zero and assigns a random helicity of $\pm1$ to each $Z$ boson. The actual non-zero average
$Z$ polarisation is correctly accounted for by applying event weights obtained with the {\sc TauSpinner}
\mbox{program~\cite{Czyczula:2012ny,Kaczmarska:2014eoa}}, which infers the most probable configuration of the initial partons
and thus the helicity of the $Z$ boson from the decay product kinematics.

\item{Kinematic filter for the decay products:}\\
If the generation of $Z\to\tau\tau$ decays were purely based on the probability distributions of the actual decay kinematics,
a large fraction of the embedded $Z\to\tau\tau$ decay products would fail the selection criteria of typical physics analyses.
In  particular the leptonic $\tau$ decays would often end up below the relevant transverse momentum thresholds.
Therefore, a kinematic $\tau$ decay filter is implemented at generator level in order to increase the effective number of
$\tau$-embedded $Z\to\mu\mu$ events entering the $\tau\tau$ selection. Instead of generating only one $\tau\tau$ decay
for each $Z$ collision data event, the {\sc Tauola} program is used to produce 1000 different kinematic configurations of
the decay products according to the appropriate probability distributions. Only the first of the 1000 decay configurations in
which the generated transverse momenta of the visible decay products ($e/\mu/\tau_{\mathrm{had}}$) exceed certain
threshold values is then selected for further processing. The thresholds can be chosen based on the final $\tau\tau$
analysis selection; for this paper, as for the $H\to\tau\tau$ lepton-hadron-channel analysis presented in Ref.~\cite{Aad:2015vsa}, 
they were set to $\pt(\tau_{\mathrm{had}})>15$~\GeV, $\pt(e)>18$~\GeV\ and $\pt(\mu)>15$~\GeV, 
i.e.~safely below the $\tau\tau$ analysis selection thresholds of $\pt(\tau_{\mathrm{had}})>20$~\GeV~and  \mbox{$\pt(e/\mu)>$ 26~\GeV}.
The selection of $\tau\tau$ decays according to these thresholds introduces kinematic biases as shown in Figure~\ref{fig:filtera}
for the visible momentum of hadronic $\tau$ decays and in Figure~\ref{fig:filterb} for the vector sum of the neutrino transverse
momenta, which corresponds to the expected missing transverse momentum in the event.
Based on all 1000 $\tau\tau$ decays generated for the given $Z$ kinematics, the probability to accept a random $\tau\tau$
decay configuration is evaluated for each event. These probabilities correspond to event-by-event filter efficiencies and are
thus propagated as weights, which correct the kinematic biases as demonstrated in Figure~\ref{fig:filter}.
\end{enumerate}
\begin{figure}[!t]
   \centering
   \subfloat[]{
       \includegraphics[width=0.45\textwidth]{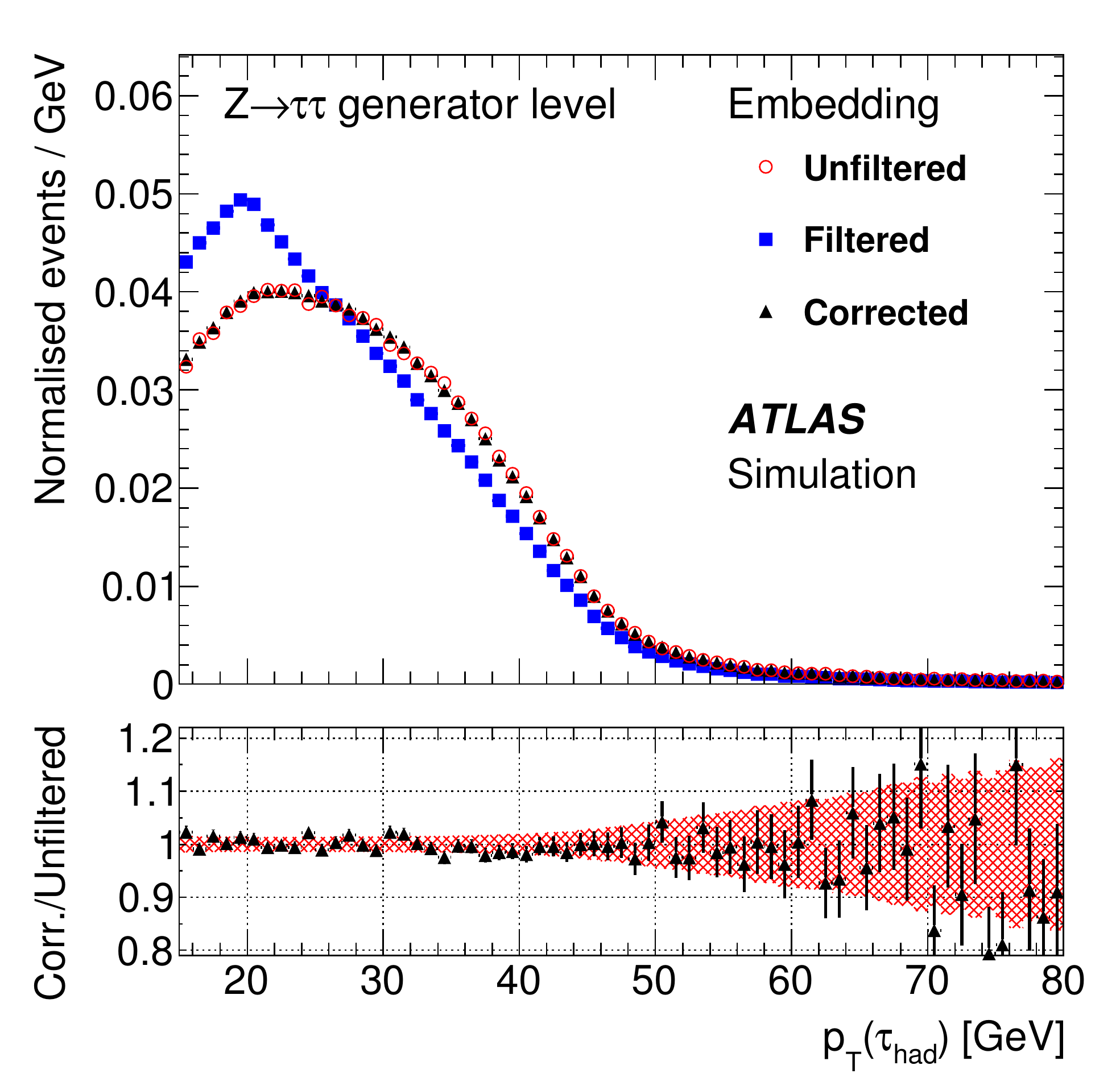}
    \label{fig:filtera}
}
\subfloat[]{
        \includegraphics[width=0.45\textwidth]{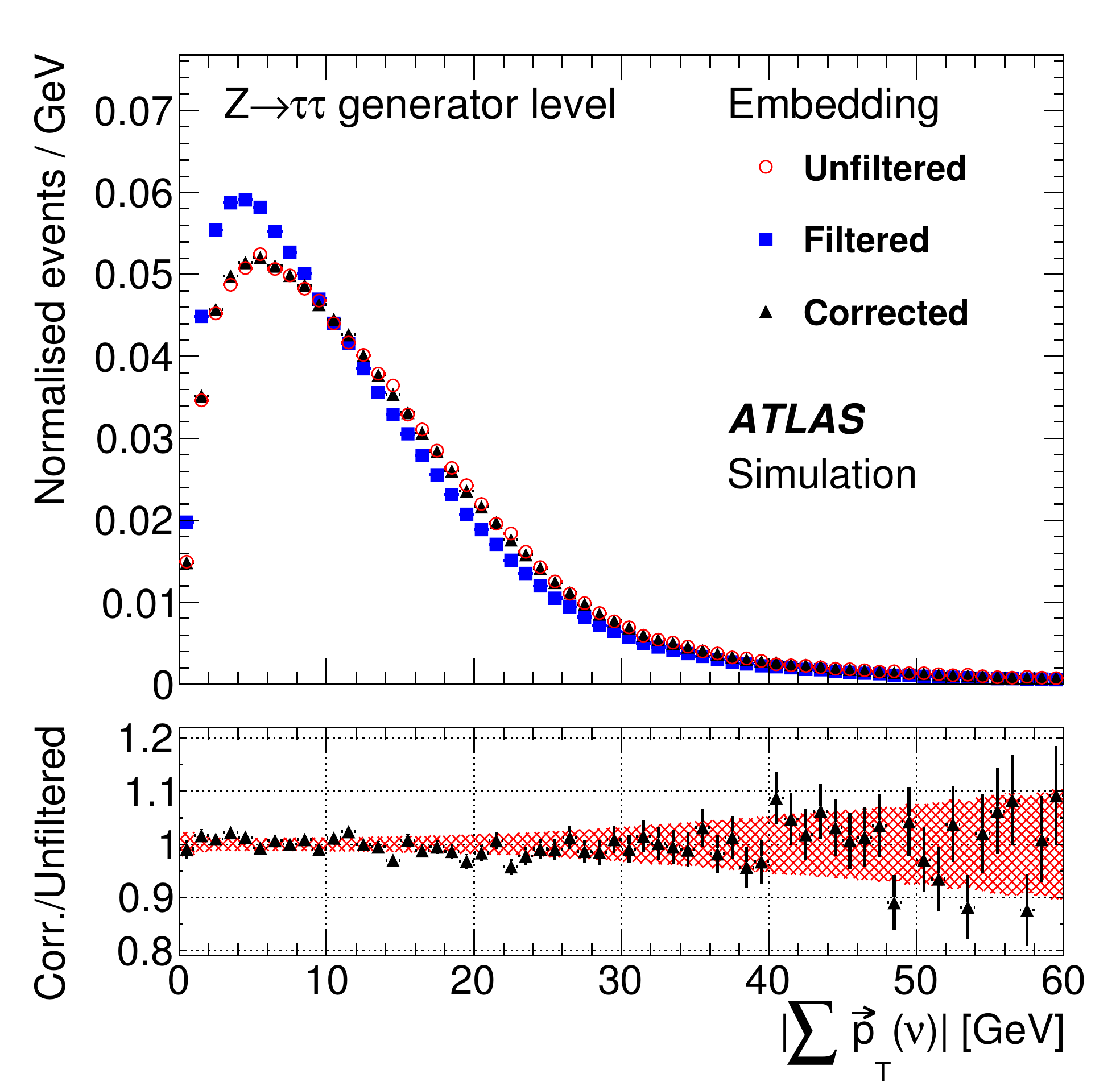}
    \label{fig:filterb}
   }
   \vspace*{-0.35cm}
     \caption{
     \label{fig:filter}
     Generator-level distributions of (a) the  $\tau_{\mathrm{had}}$ transverse momentum
and (b)  the summed transverse momenta of all neutrinos for $\tau$-embedded events without filter (red open circles),
after applying the filter (blue squares) and after applying the filter with filter weights (black triangles) as described in the text.
The lower panels show the relative deviation of the corrected distributions from the unfiltered ones. The red shaded error
band and the black error bars correspond to the statistical uncertainty from the unfiltered and filtered events, respectively.
}
 \vspace*{-0.35cm}
\end{figure}
\item{Detector simulation of the $Z\to\tau\tau$ decay:}\\
The result corresponds to a standard event generator output for a $Z\to\tau\tau$ decay without any underlying-event effects,
but otherwise based on the standard ATLAS MC configuration~\cite{Aad:2010ah}, which is then handed over to the full
ATLAS detector simulation and reconstruction. In order to avoid double counting in the later merging with the corresponding
collision data event, the calorimeter noise is switched off during the simulation.
In the following, the output of this simulation step is referred to as a \emph{mini event}.
\end{enumerate}

\begin{figure}[!h]
\vspace*{-0.26cm}
   \centering
    \subfloat[]{
       \includegraphics[width=0.41\textwidth]{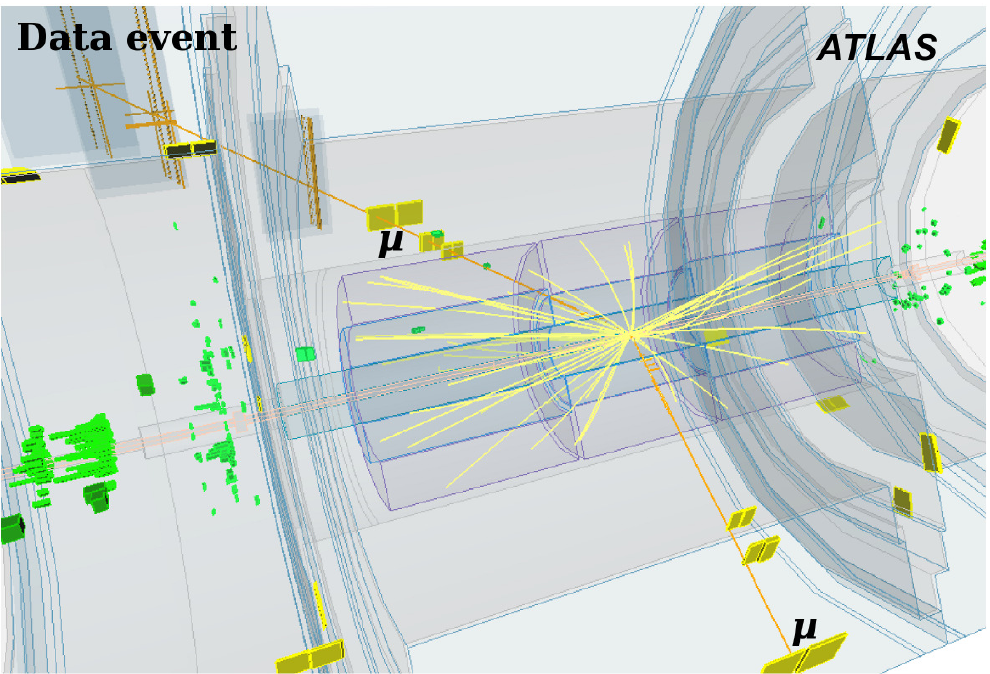}
       \label{fig:evtdispla}
}
\subfloat[]{
              \includegraphics[width=0.41\textwidth]{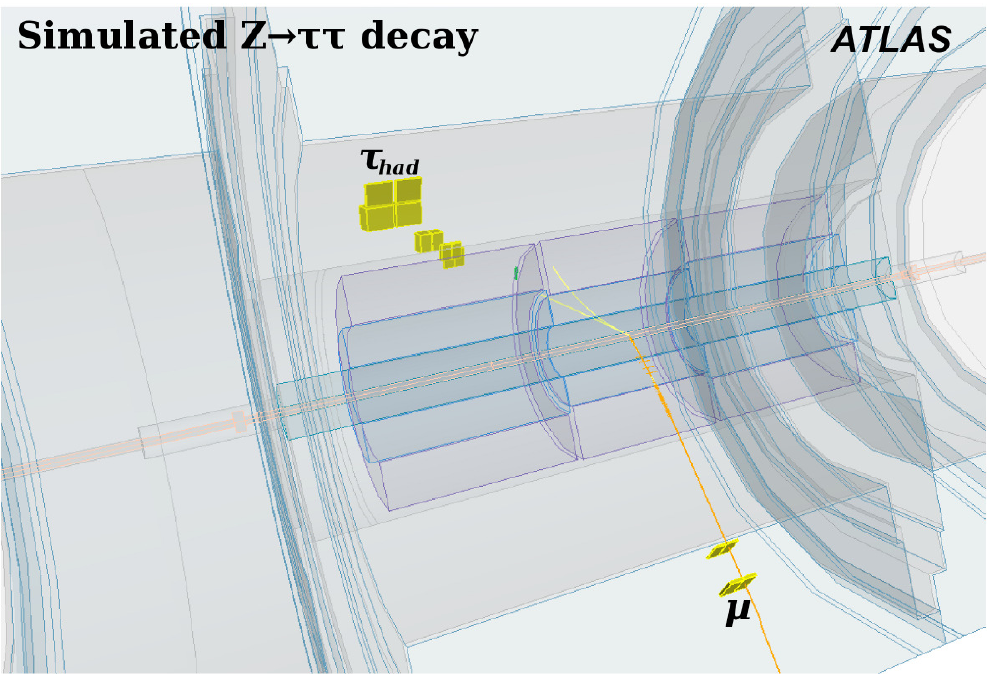}
              \label{fig:evtdisplb}
}
\\
\subfloat[]{
              \includegraphics[width=0.41\textwidth]{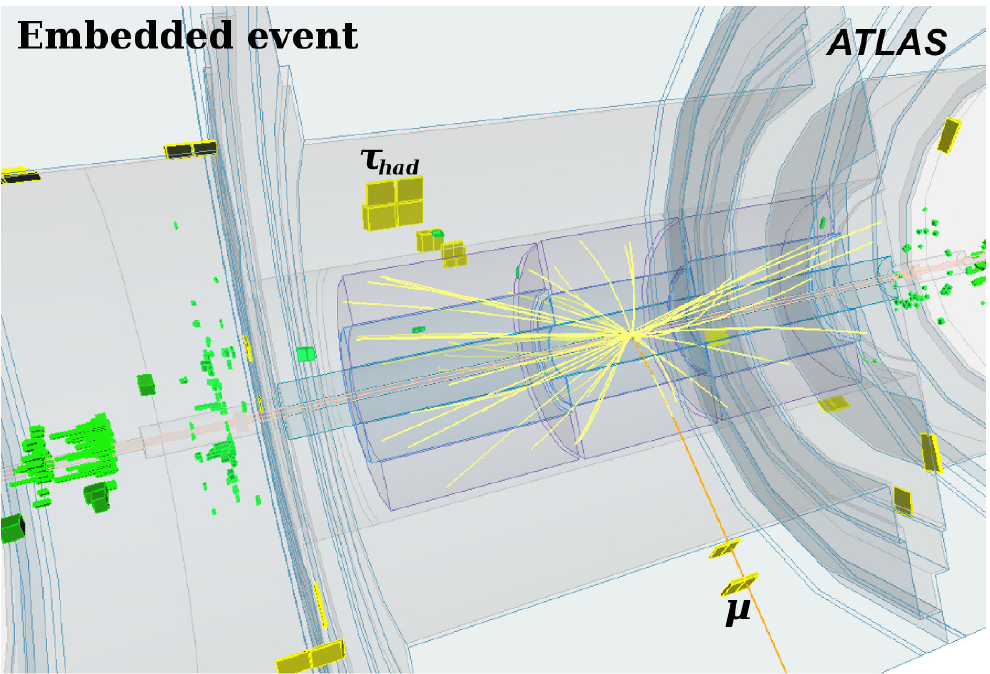}
              \label{fig:evtdisplc}
}
\vspace*{-0.3cm}
 \caption{
Displays of (a) a  $Z\rightarrow\mu\mu$ candidate event selected from the collision data,
(b) the corresponding simulated $Z\rightarrow\tau\tau$ mini event and (c) embedded hybrid event.
Here, one of the $\tau$ leptons decays into a muon and the other one hadronically.
\label{fig:evtdispl}
}
\vspace*{-0.4cm}
\end{figure}

\begin{enumerate} \addtocounter{enumi}{3}
\item{Merging of data and simulated event:}\\
In order to replace the muons in the selected $Z\rightarrow\mu\mu$ data events with the corresponding simulated $\tau$
leptons, all tracks associated with the original muons are removed from the data event.
The calorimeter cells associated with the muons are subtracted according to the following procedure: a $Z\rightarrow\mu\mu$
decay with the same kinematics as the original event (and without the underlying event or pile-up interactions) is simulated.
The calorimeter cell energies in the simulated event are subtracted from the data event.

All calorimeter cell energies from the simulated mini event are then added to the corresponding data cell energies,
and all tracks are copied into the corresponding event. This inserts the pure $Z\rightarrow\tau\tau$ decay into the data
environment, keeping the event properties as close to data conditions as possible.
\item{Reconstruction of the embedded events:}\\
Starting from the modified cell energies and the merged set of tracks, the hybrid $Z\rightarrow\tau\tau$ events are submitted to the  ATLAS event reconstruction for collision data, which recreates the complete physics object final 
state by re-running all standard event reconstruction algorithms except for the track reconstruction.

\end{enumerate}
The procedure is further illustrated by Figure~\ref{fig:evtdispl}, showing example displays of a
$Z\rightarrow\mu\mu$ input event,  a correspondingly simulated $Z\rightarrow\tau\tau$
mini event (with one $\tau$ lepton decaying into a muon and the other one hadronically), and the resulting
embedded hybrid event.

\subsection{Special properties of the $\tau$-embedded event samples}
\label{subsec:corr}
While in most respects the $\tau$-embedded samples can be treated within physics analyses as
standard collision data, there are a few special properties to be considered:
\begin{itemize}
\item The $Z\to\mu\mu$ input data are subject to trigger and offline selection efficiencies, which particularly affect analyses
with low $\pt$ selection thresholds for the $\tau$ decay products. To account for these efficiencies, correction factors as a
function of the transverse momenta and pseudorapidities of the input muons are extracted according
to Refs.~\cite{TriggerPaper,Aad:2014rra} and applied to the $\tau$-embedded samples.

\item As discussed in Section~\ref{subsec:proc}, instead of recreating the charged-particle tracks from the tracking
detector hits, the embedding procedure is performed with reconstructed tracks. As a side effect, the trigger response
for the $\tau$-embedded events is not available, since it would require the hit-level information. Therefore, any effect of
the analysis-specific trigger selection needs to be evaluated and corrected for, e.g. through a parameterisation of the
trigger efficiency measured in data. For the validation in Section~\ref{subsec:tauval}, such corrections were derived
corresponding to the $Z\to\tau\tau$  selection described in Section~\ref{subsec:selection} and applied to the
$\tau$-embedded samples.

\item The selected $Z\to\mu\mu$ input data sample is of high purity, but small contaminations from other processes, e.g.
$t\bar{t}$ production, might be enhanced to relevant levels by selection requirements applied during physics analyses.
Double counting of these contributions must hence be avoided when combining the $\tau$-embedded events with
other samples to construct a complete background model. In recent analyses, e.g. in Ref.~\cite{Aad:2015vsa}, this is
achieved by rejecting events from simulated samples of other background processes if they produce two $\tau$ leptons that
fulfil the kinematic $Z\to\mu\mu$ input selection at generator level. The corresponding $\tau\tau$ final states are already
included in the $\tau$-embedded sample as obtained from the corresponding $\mu\mu$ background contamination from
other processes.

\item In deriving the kinematics of the embedded $\tau$ leptons from the reconstructed muons selected from the ATLAS data,
the true kinematics of the $Z$ decay are folded with the resolution of the muon reconstruction. Final-state radiation (FSR) from
the input muons can also modify the kinematics of the embedded objects. Both effects are unavoidable and inseparable in the
embedding of data events, but they can be studied separately using simulated samples and are found to be small
(cf.~Section~\ref{subsec:tauval}).

\item
While the $\tau$-embedded samples constitute a largely data-driven model of $Z\to\tau\tau$ events, the $\tau$ leptons and
their decay products are based on simulation, and systematic uncertainties associated with the MC description of
$\tau$ decays and the corresponding detector response need to be considered within physics analyses. Further
documentation of these systematic uncertainties, e.g. for the hadronic $\tau$ decays, can be found in Ref.~\cite{Aad:2014rga}.
\item The size of the $\tau$-embedded samples is naturally limited by the available number of $Z\rightarrow\mu\mu$ data
events. Compared to a corresponding selection of $\tau\tau$ final states from the data, this number is effectively enhanced by
applying the kinematic filter described in Section~\ref{subsec:proc}.
\end{itemize}

\subsection{Systematic uncertainties}
\label{subsec:syst}
%

%
Two different sources of systematic uncertainty are considered, which are motivated by the technical implementation of
the embedding method and are thus estimated from the following variations of the embedding procedure:
\begin{figure}[!b]
\vspace*{-0.7cm}
   \centering
   \subfloat[]{
       \includegraphics[width=0.45\textwidth]{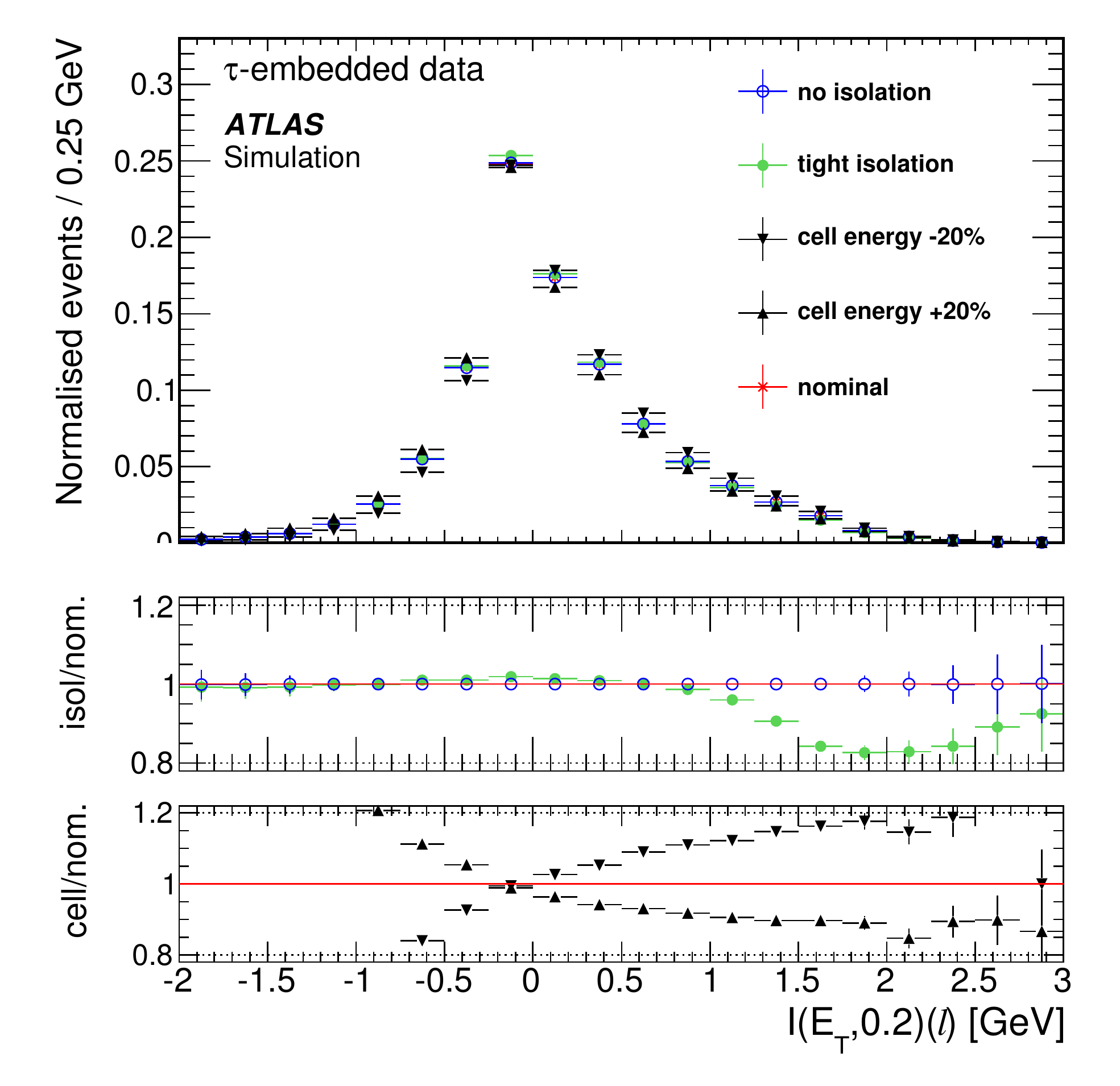}
       \label{fig:systa}
    }
    \subfloat[]{
       \includegraphics[width=0.45\textwidth]{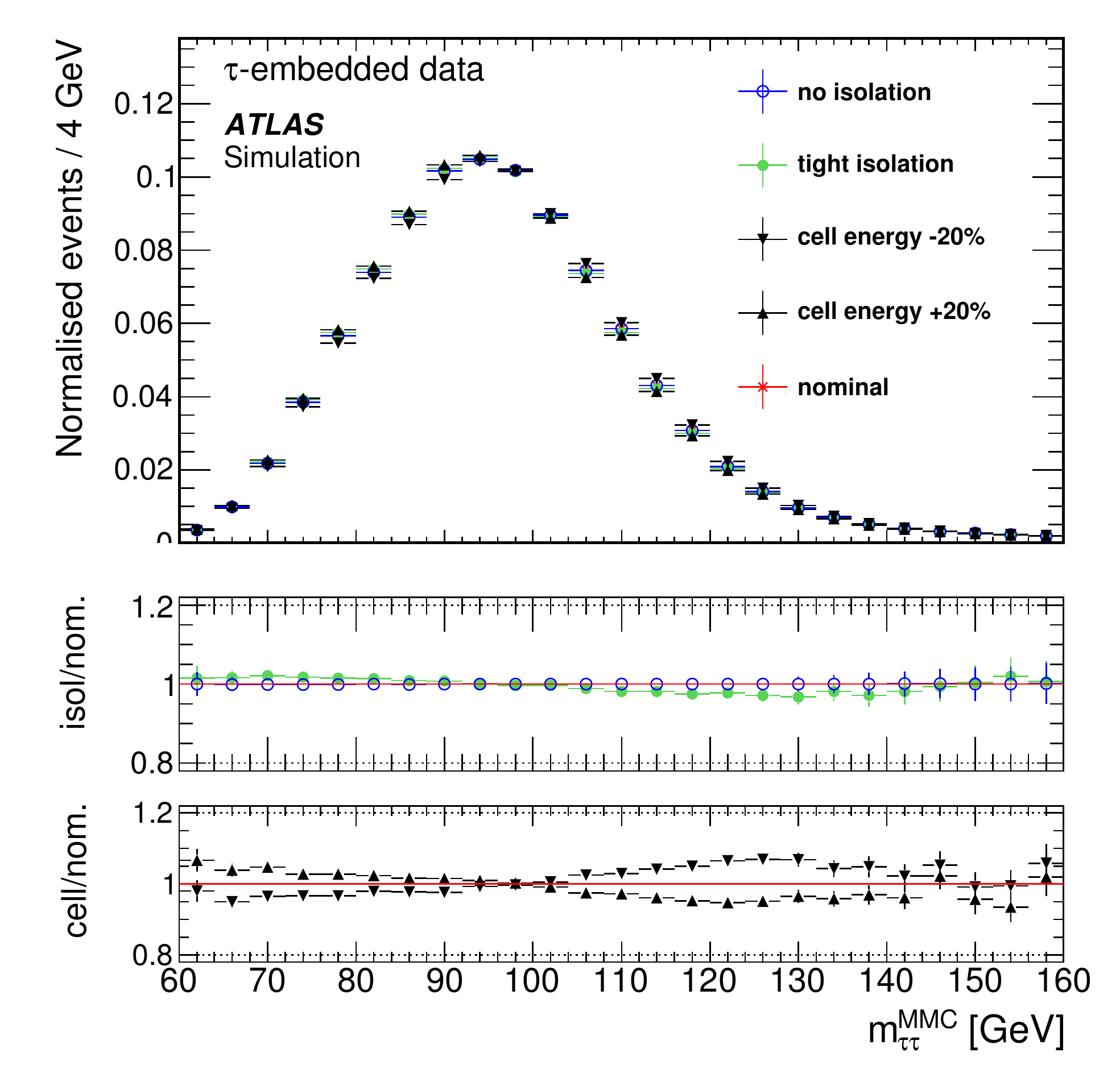}
       \label{fig:systb}
    }
\vspace*{-0.2cm}    
\caption{
\label{fig:syst}
Distributions of (a) the calorimeter isolation of the selected lepton and (b) the $\tau\tau$ invariant mass obtained with the MMC,
illustrating the effects of systematic variations as described in the text: scaling the subtracted cell energy by $\pm20\%$ and
applying tight / no isolation requirements in the $Z\rightarrow\mu\mu$ selection. The ratios of the distributions before and after
specific systematic variations are included as well: the upper ratio plot shows the effect of no (tight) isolation in blue (green), in
the lower one the effect of scaling the subtracted cell energy by +20\% (-20\%) is illustrated by triangles pointing upwards
(downwards). In both plots the red lines correspond to the nominal embedded sample.
}
\end{figure}

\begin{enumerate}
\item
The isolation requirement applied in the selection of the $Z\to\mu\mu$ input events can affect the environment of the
embedded objects in the final event. It is thus varied in two alternative selections: the nominal isolation criterion of
$I(\pt,0.4)/\pt(\mu)<0.2$ is either completely removed or tightened to $I(\pt,0.4)/\pt(\mu)<0.06$ and
$I(E_{\rm T},0.2)/\pt(\mu)<0.04$. These variations mainly affect the properties of the embedded objects, but they
additionally provide an estimate of the background contamination from $\mu\mu$ final states with non-prompt muons in
the $\tau$-embedded samples.
\item
The subtraction of cell energy associated with the muon is based on the simulated calorimeter response, which can be
subject to large uncertainties. Therefore, the simulated energy in each cell is scaled by $\pm20\%$ before the subtraction
from the data event. The size of this variation was motivated by the results of comparisons of $\tau$-embedded
collision-data and simulated events to standard \mbox{$Z\to\tau\tau$} MC samples.
\end{enumerate}
For all embedded event samples listed in Section~\ref{subsec:samples}, the different variations are produced in parallel.
The resulting datasets are then used to derive and validate the embedding-related systematic uncertainties. Different
selection efficiencies, e.g. due to the modified isolation requirements, are absorbed by normalising the systematic variations
to the default sample. For both estimates of systematic uncertainties, the remaining shape uncertainties are later symmetrised
to the larger of the two variations, in particular compensating for the non-symmetric isolation criteria. Figure~\ref{fig:syst}
illustrates the effect on the distributions of two example quantities after the $Z\to\tau\tau$ selection as described in
Section~\ref{subsec:selection}.

Modifications of the input muon kinematics due to final-state radiation or the detector resolution, which could be considered as
somewhat more fundamental sources of systematic effects, do not directly enter the above definitions of embedding-related
uncertainties. Their impact is, however,  expected to be correlated with the variations of the cell energy subtraction and the
muon isolation and in fact turns out to be small in comparison, as demonstrated in Section~\ref{subsec:tauval}.
\clearpage
\section{Validation}
\label{sec:validation}
%
A careful validation of the embedding procedure is performed based on different combinations of
the event samples described in Section~\ref{subsec:samples}. The results of these studies
are discussed in the following.  All distributions are normalised to unit area unless stated otherwise.
%
\subsection{$Z\to\mu\mu$-based validation}
\label{subsec:muval}
%
The first set of studies is based on muon-embedded data and MC samples, where the original muons are removed
and replaced with the decay products of correspondingly simulated $Z\to\mu\mu$ decays.
\begin{figure}[!b]
\vspace*{-4mm}
   \centering
   \subfloat[]{
        \includegraphics[width=0.5\textwidth]{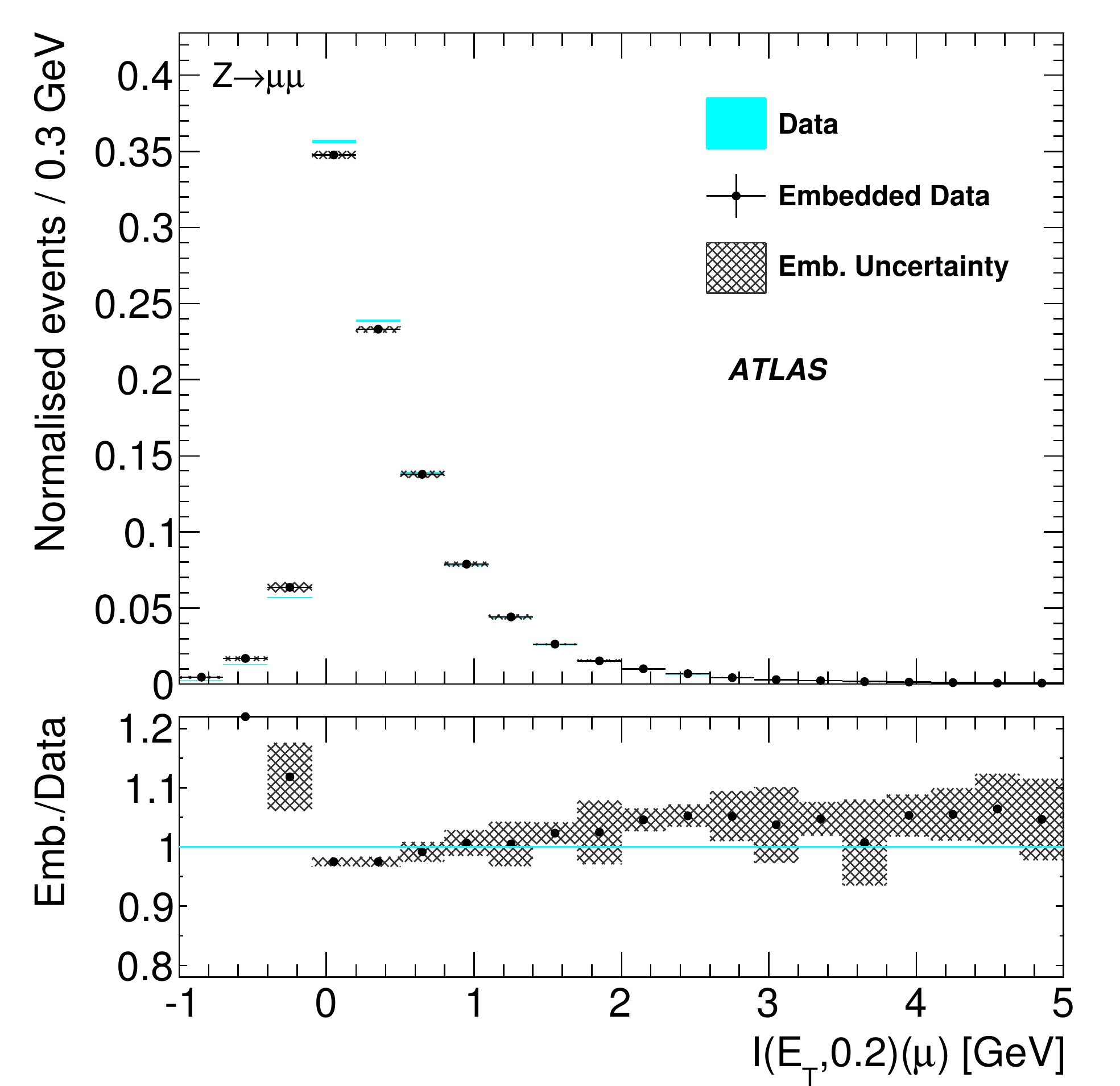}
        \label{fig:mureco1a}
}
\subfloat[]{
        \includegraphics[width=0.5\textwidth]{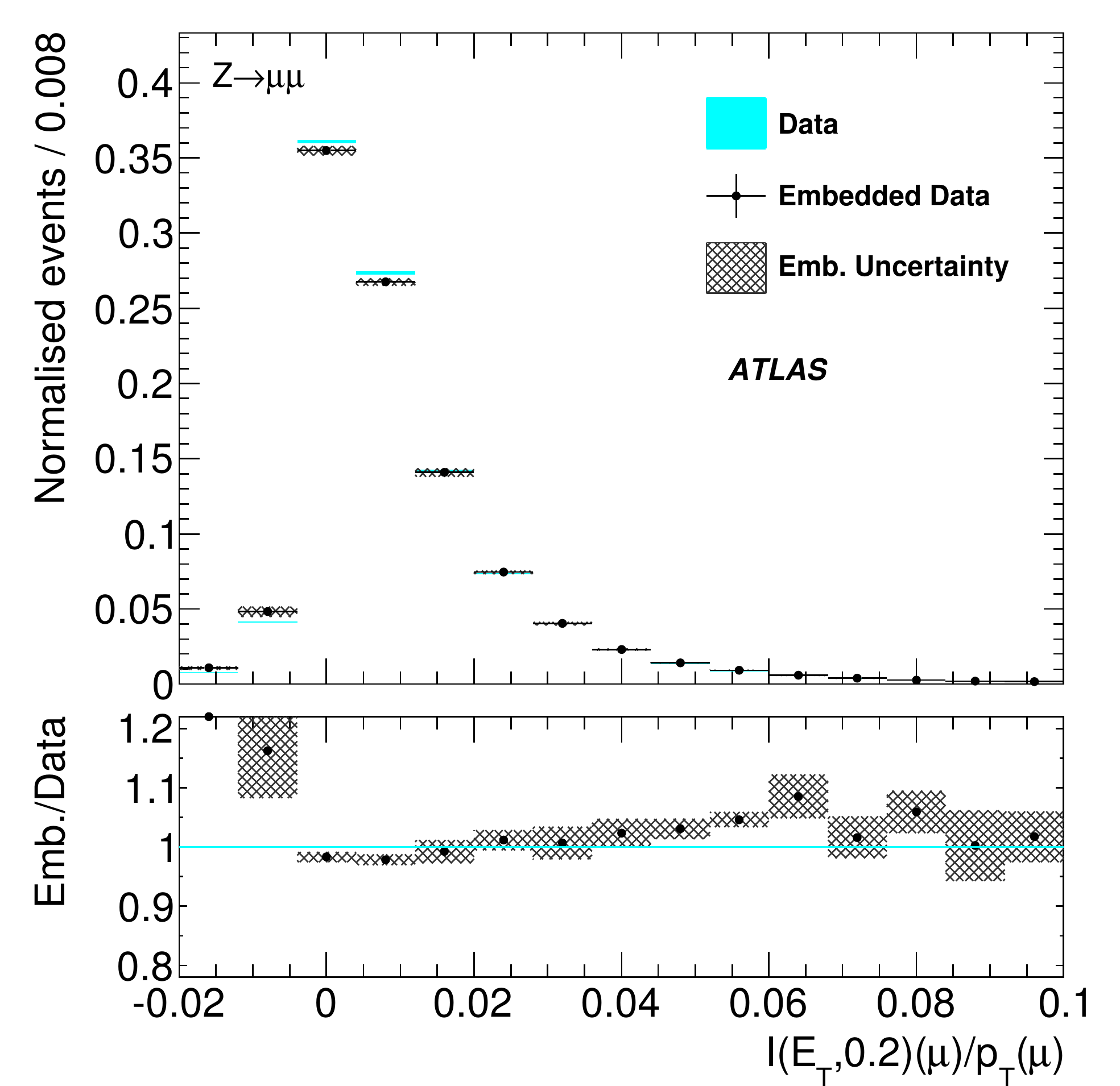}
        \label{fig:mureco1b}
}
\caption{
\label{fig:mureco1}
Comparison of $Z\rightarrow\mu\mu$ data events before (blue) and after $\mu$ embedding (black points) in terms of (a)
the calorimeter isolation and (b) the relative calorimeter isolation in a cone $\Delta R=0.2$, each including ratios showing
the relative differences of the distributions after $\mu$ embedding. The grey hatched error band corresponds to the
cell energy systematic uncertainties of the $\mu$-embedded events, as described in Section~\ref{subsec:syst}.
}
\end{figure}
In this case, events with $Z\to\mu\mu$ decays and jets constitute both the input and the output samples and thus
distributions of any quantity for the same events before and after the embedding can be compared directly.
Such comparisons provide a powerful validation of most aspects of the procedure by testing for biases introduced in
the removal of tracks and cells associated with the input muons, the stand-alone simulation of the $Z$ mini event
or the creation and re-reconstruction of the embedded hybrid event. None of the trigger and selection efficiency
corrections discussed in Section~\ref{subsec:corr}
are applied here.

In order to investigate possible distortions of the detector response close to the input muons, Figure~\ref{fig:mureco1}
compares the distributions of the absolute ($I(E_{\rm T},0.2)$) and relative ($I(E_{\rm T},0.2)/\pt$) muon calorimeter
isolation as defined in Section~\ref{subsec:selection}, before and after $\mu$ embedding. Here, the displayed errors
do not include the isolation systematic uncertainty, which is obtained by varying an explicit cut on
the relative calorimeter isolation, cf.~Section~\ref{subsec:syst}, and is thus not well defined in these specific comparisons.
The observed changes in the distributions, which indicate fluctuations in the estimation of the calorimeter energy associated
to the input muons based on an independent simulation discussed in~Section~\ref{subsec:proc}, are not fully
covered by the remaining embedding-specific uncertainties. However, this mainly concerns negative isolation values, which
are far away from standard isolation requirements as also used for the studies presented in this paper, and the region with
$I(E_{\rm T},0.2)/\pt>0.04$, where the undisplayed isolation uncertainty becomes very large by construction.
In corresponding comparisons, the kinematics of additional jets in the event are
found to be unaffected by the embedding procedure.

\begin{figure}[!t]
   \centering
\subfloat[]{
       \includegraphics[width=0.5\textwidth]{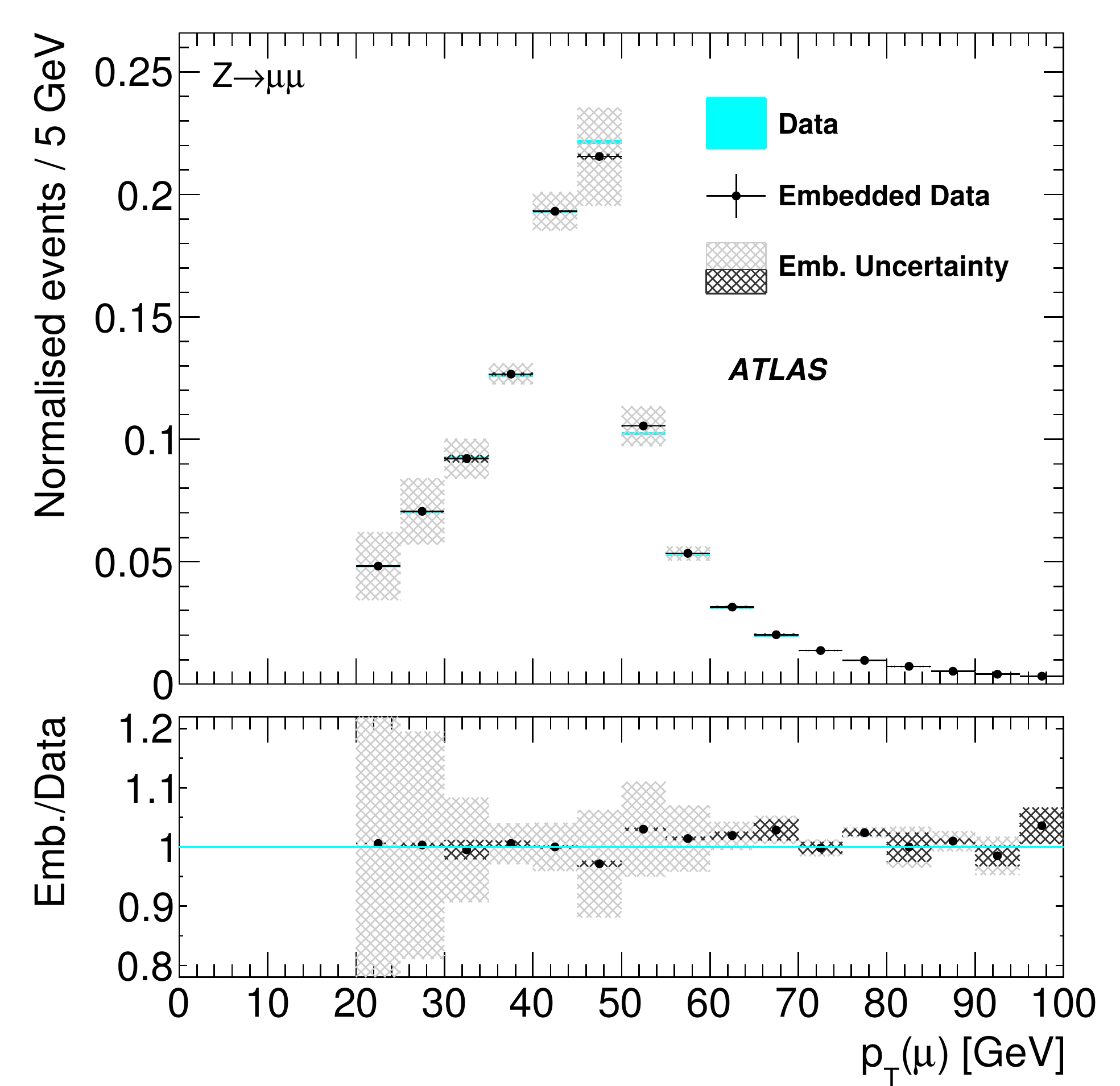}
       \label{fig:mureco3a}
}
\subfloat[]{
       \includegraphics[width=0.5\textwidth]{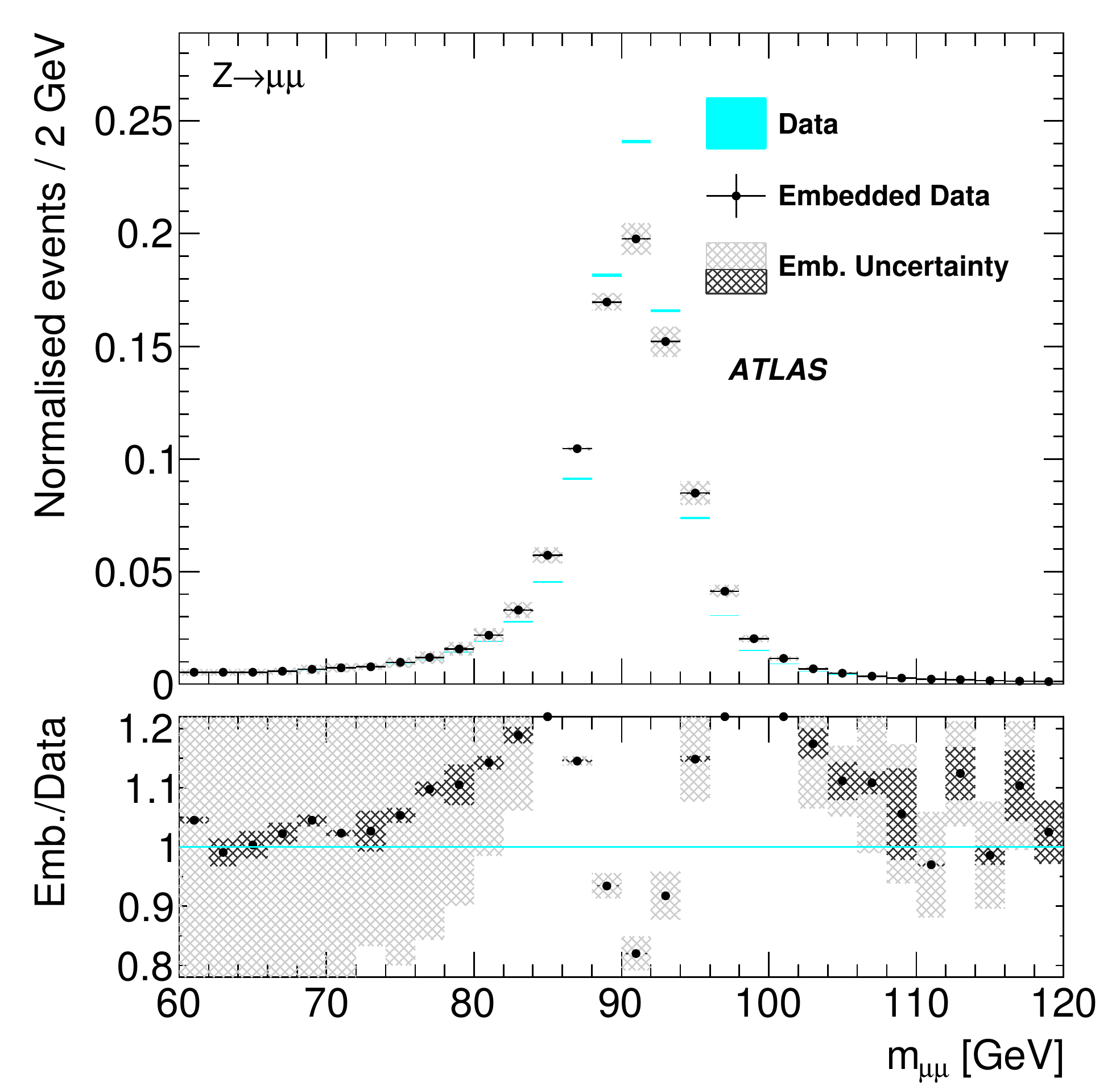}
       \label{fig:mureco3b}
}
\caption{
\label{fig:mureco3}
Comparison of  $Z\rightarrow\mu\mu$ data events before (blue) and after $\mu$ embedding (black points):
(a) transverse momentum of the leading muon and (b) dimuon mass, each including ratios showing the relative
differences of the distributions after $\mu$ embedding. The light (dark) grey hatched error band corresponds to
the sum in quadrature of cell + isolation (cell only) systematic uncertainties of the $\mu$-embedded events.
}
\end{figure}

\begin{figure}[!t]
   \centering
\subfloat[]{
       \includegraphics[width=0.50\textwidth]{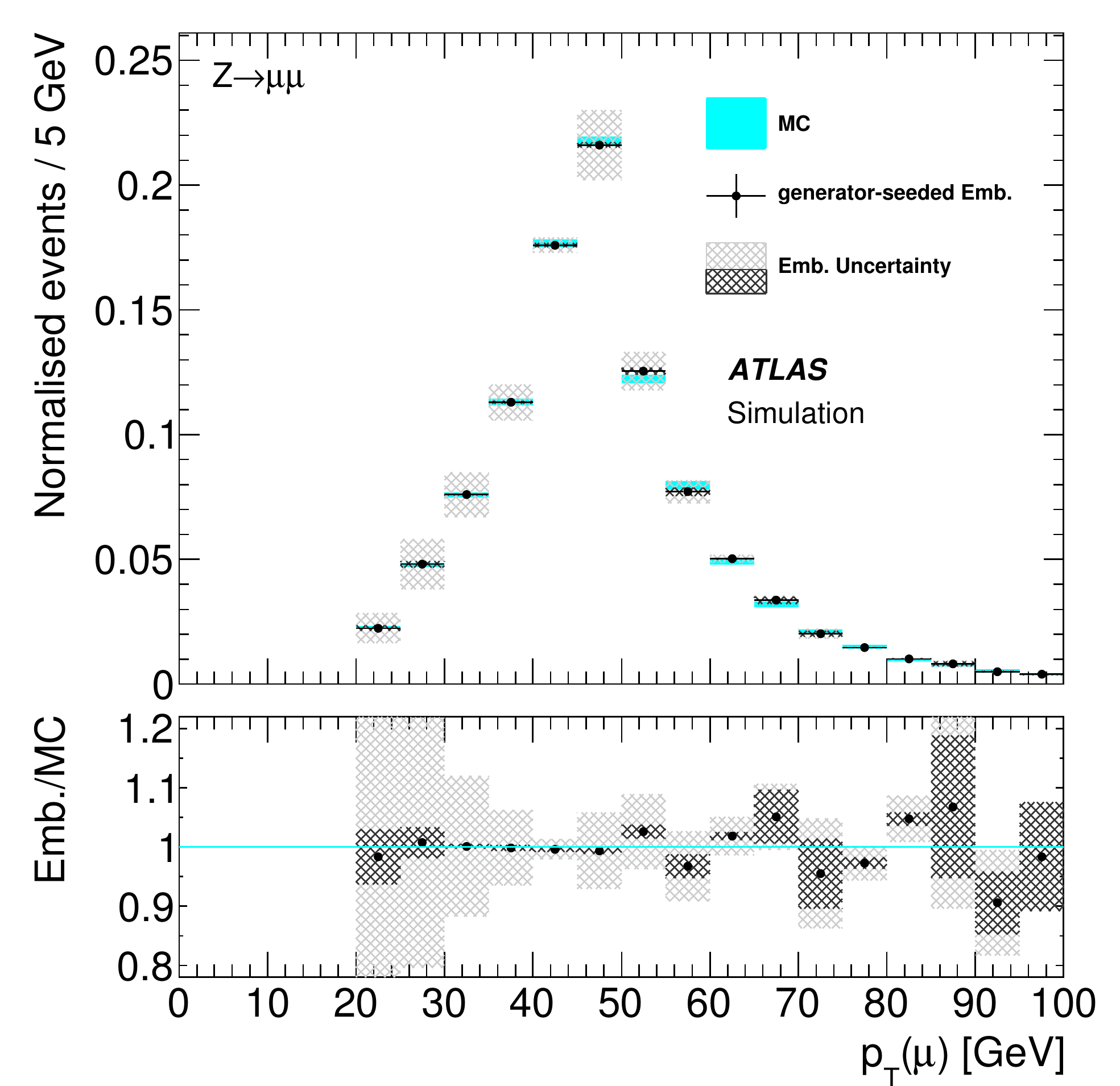}
       \label{fig:mureco4a}
}
 \subfloat[]{
       \includegraphics[width=0.50\textwidth]{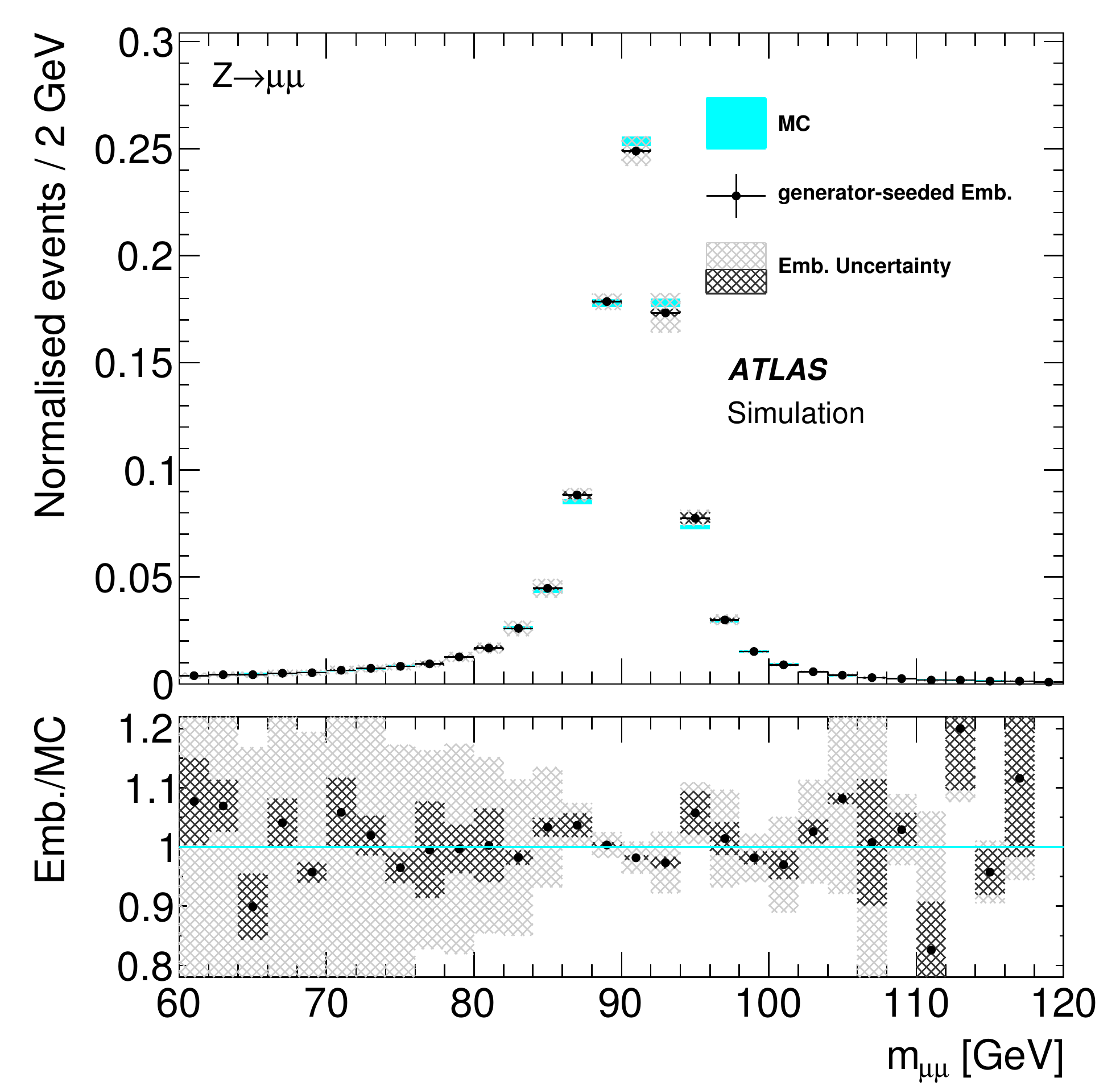}
       \label{fig:mureco4b}
}
\caption{
\label{fig:mureco4}
Comparison of  $Z\rightarrow\mu\mu$ MC events (blue) and generator-seeded $\mu$ embedding (black points):
(a) transverse momentum of the leading muon and  (b) dimuon mass, each including ratios showing the relative
differences of the distributions after generator-seeded $\mu$ embedding.
The light (dark) grey hatched error band corresponds to the sum in quadrature of cell + isolation (cell only) systematic
uncertainties and the statistical uncertainties of the $\mu$-embedded events.
}
\end{figure}

For quantities directly related to the muon four-momenta, most changes are found to be within the uncertainties;
for example, Figure~\ref{fig:mureco3}\subref{fig:mureco3a} shows  the transverse momentum of the leading muon. In
some cases, however, larger effects are observed, in particular for the dimuon invariant mass as depicted in
Figure~\ref{fig:mureco3}\subref{fig:mureco3b}; small differences are also found at the low end of the distributions of
the transverse momentum of the dimuon system and of the missing transverse momentum.
Such differences are actually expected since the kinematics of the embedded events are based on reconstructed input
muons and thus are potentially modified by the detector resolution and final-state radiation (FSR),
as explained in Section~\ref{subsec:corr}.
This is investigated further by using generator-seeded embedded samples, where simulated
$Z\to\mu\mu$ events are used as input and the kinematics of the embedded objects is derived
from the generator-level muon momenta instead of the reconstructed information,
cf.~Section~\ref{subsec:samples}, thus removing FSR and muon reconstruction effects.
This indeed improves the agreement in the muon-related distributions shown  for the leading muon $\pt$
and the dimuon mass in Figure~\ref{fig:mureco4}.
While these simulation-based studies confirm the source of the differences in Figure~\ref{fig:mureco3}, muon reconstruction
and FSR effects unavoidably enter the embedding of data events. For the eventual applications of $\tau$ embedding, however,
these differences turn out to be negligible as demonstrated in the next section.
\newpage

\subsection{$Z\to\tau\tau$-based validation}
\label{subsec:tauval}
%

The $Z\to\mu\mu$-based results presented above already provide confidence that the technical implementation of the
embedding procedure is working correctly. Nevertheless, direct comparisons of $\tau\tau$ final states must also be performed
in order to conclusively validate the modelling of $Z\to\tau\tau$ events provided by the final $\tau$-embedded samples.
Since it is difficult to obtain a sufficiently pure $Z\to\tau\tau$ reference sample from the collision data, the validation is
mainly based on comparisons of $\tau$-embedded $Z\to\mu\mu$ MC events to standard $Z\to\tau\tau$ MC samples.
Still, comparisons of selected $Z\to\tau\tau$ collision data to a combined background model including $\tau$-embedded data
are also provided in the last part of this section.

\subsubsection*{Input muon radiation and reconstruction effects}
The embedding procedure includes two effects related to the input muons that are unavoidable by construction:
\begin{figure}[!t]
   \centering
      \subfloat[]{
        \includegraphics[width=0.5\textwidth]{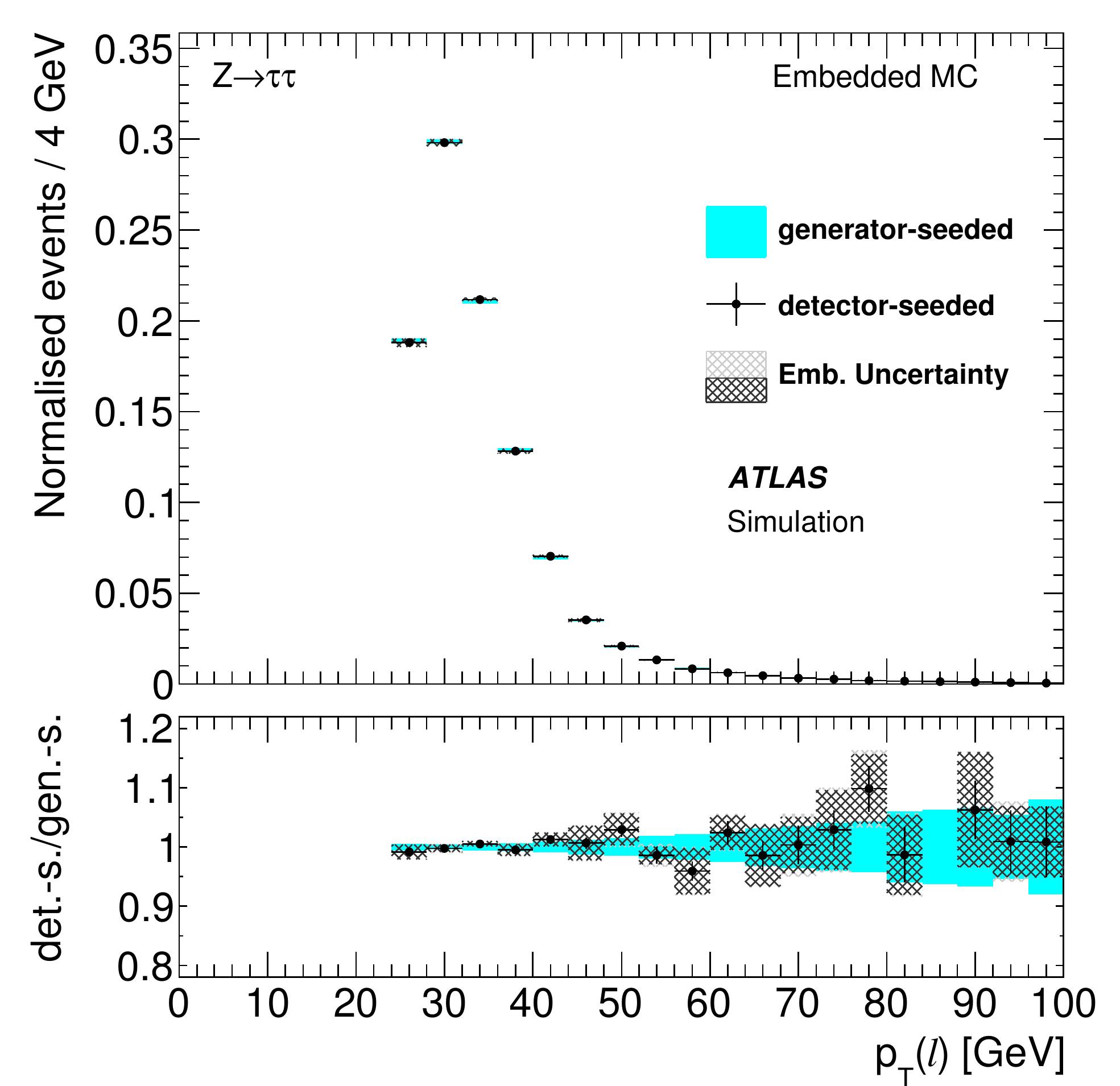}
           \label{fig:taurecotrutha}
   }
   \subfloat[]{
        \includegraphics[width=0.5\textwidth]{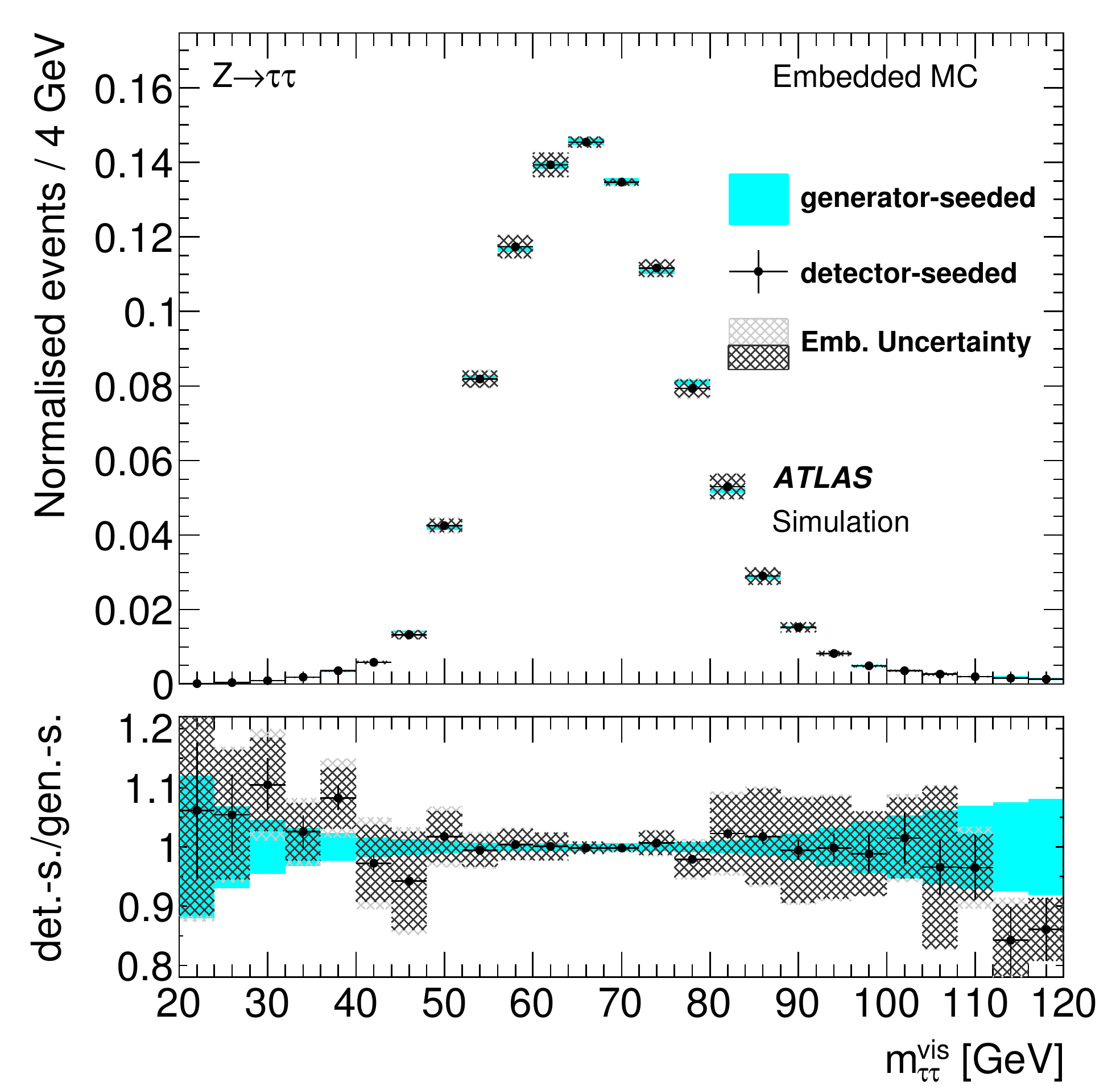}
           \label{fig:taurecotruthb}
   }
     \caption{
     \label{fig:taurecotruth}
 Comparison of generator-seeded (gen.-s.), in blue, and detector-seeded (det.-s.), as black points, $\tau$-embedded
 $Z\rightarrow\mu\mu$ MC events: (a) transverse momentum of the leading lepton and
(b) invariant mass of the visible $\tau\tau$ decay products, each including ratios showing the relative differences of
the distributions from detector-seeded $\tau$ embedding. The blue error band in the ratio plots corresponds to the
statistical uncertainties of the generator-seeded events, and the black error bars are the statistical uncertainties associated
with the detector-seeded embedded events. The light (dark) grey hatched error band corresponds to the sum in quadrature
of cell + isolation (cell only) systematic uncertainties and the statistical uncertainties of the detector-seeded
$\tau$-embedded events.
}
\end{figure}
the resolution of the reconstructed muon momenta used to derive the kinematics of the embedded mini event and
FSR from the input muons.
In order to judge if the resolution effects observed in Section~\ref{subsec:muval} are significant for the eventual $\tau$
embedding, Figure~\ref{fig:taurecotruth} compares the distributions of the $\tau$ decay lepton transverse momentum
and of the invariant mass of the visible $\tau\tau$ decay products, $m^{\mathrm{vis}}_{\tau\tau}$, for generator- and
detector-seeded $\tau$ embedding. These comparisons demonstrate that the uncorrected resolution and final-state radiation
of the input muons are negligible in the case of reconstructed $\tau\tau$ final states, for which the mass resolution is
dominated by the neutrinos produced in the $\tau$ decay.

\subsubsection*{Comparison of $\tau$-embedded $Z\to\mu\mu$ MC samples with standard $Z\to\tau\tau$ MC}
\begin{figure}[!t]
   \centering
\subfloat[]{
       \includegraphics[width=0.50\textwidth]{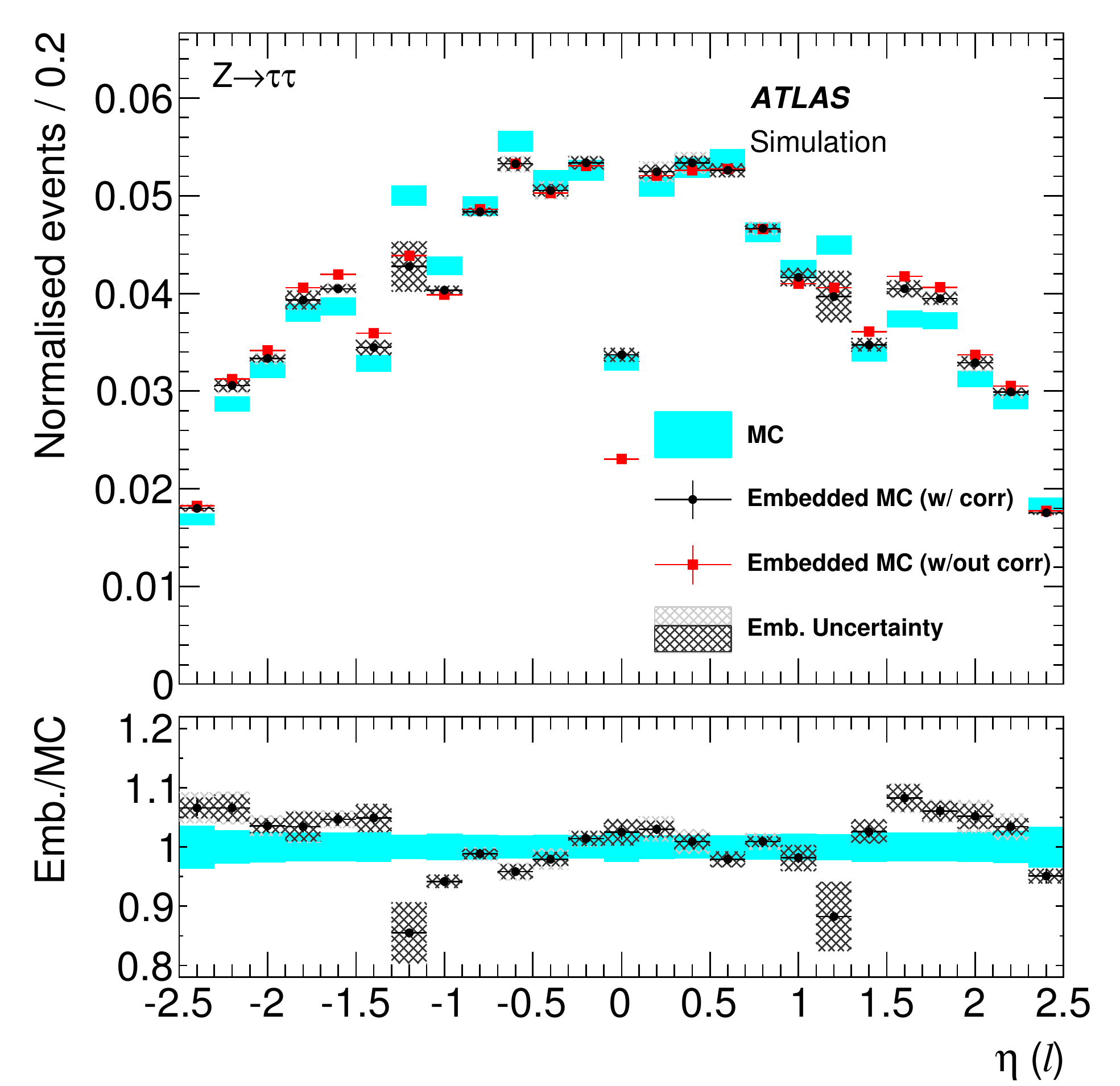}
       \label{fig:taucorr1a}
}
 \subfloat[]{
       \includegraphics[width=0.50\textwidth]{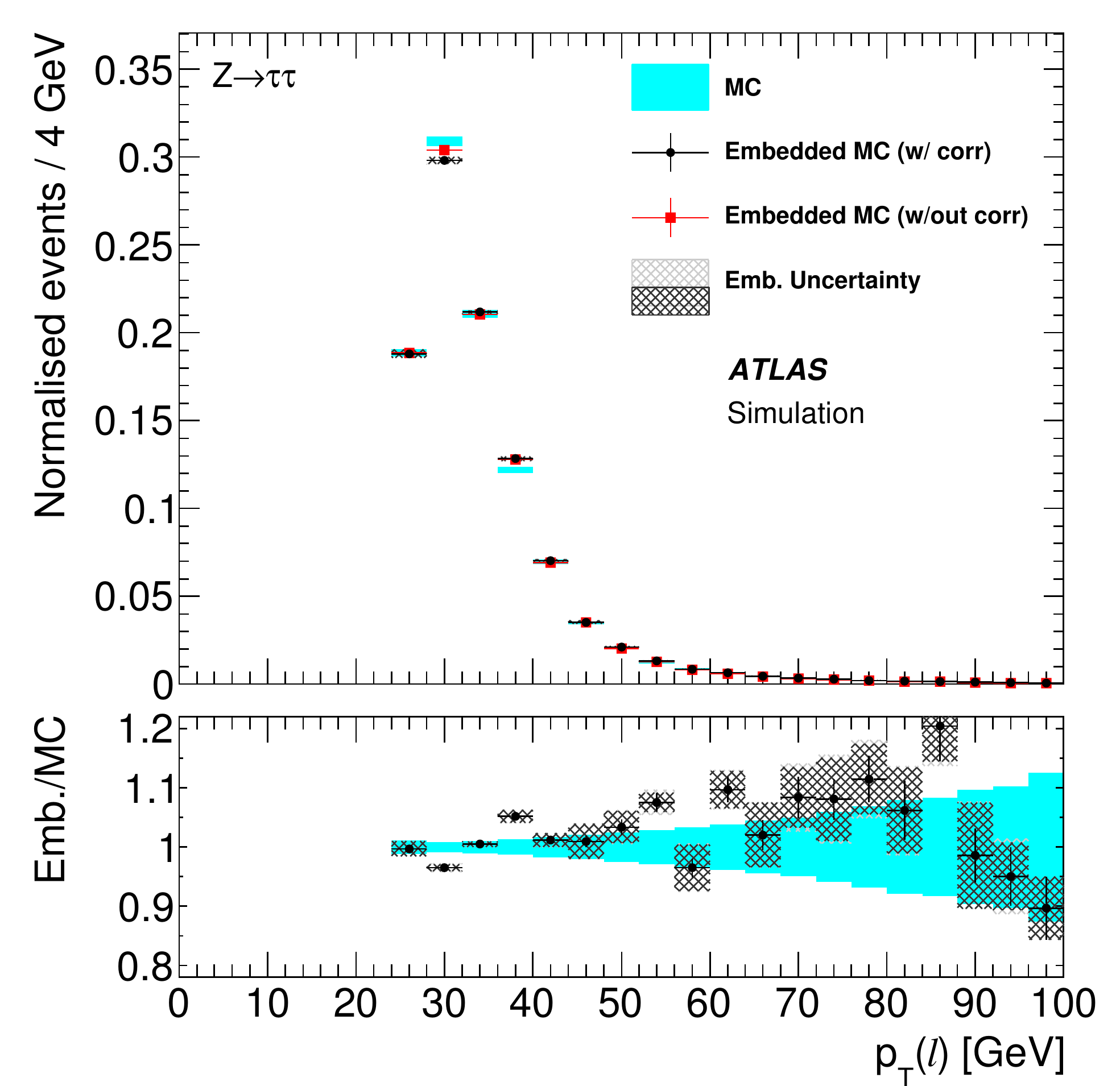}
       \label{fig:taucorr1b}
}
\caption{
\label{fig:taucorr1}
Comparison of $\tau$-embedded $Z\rightarrow\mu\mu$ MC events (black points) with $Z\rightarrow\tau\tau$ MC events
(blue) for (a) the pseudorapidity and  (b) the transverse momentum of the $\tau$ decay lepton, each including ratios
showing the relative differences of the $\tau$-embedded distributions. In addition, the red squares show the
distributions obtained from the $\tau$-embedded MC sample before applying the embedding-specific corrections.
The blue error band in the ratio plots corresponds to the statistical uncertainties of the $Z\rightarrow\tau\tau$ MC sample.
The black error bars are the statistical uncertainties associated with the corrected $\tau$-embedded events. The light (dark)
grey hatched error band corresponds to the sum in quadrature of cell + isolation (cell only) systematic uncertainties and
the statistical uncertainties of the corrected $\tau$-embedded events.
}
\end{figure}

In contrast to a data-to-data comparison of $\tau\tau$ final states, which necessarily includes contaminations from other
background processes, the $\tau$ embedding of simulated $Z\to\mu\mu$ events and subsequent comparison to standard
$Z\to\tau\tau$ MC samples provides a well-defined way to further study the method at the $\tau\tau$ level. Here, as
opposed to the studies presented in Section~\ref{subsec:muval}, the two compared distributions are obtained from
statistically independent event samples.
Also, the corrections discussed in Section~\ref{subsec:corr}, including those related to the selection of
the $Z\to\mu\mu$ events used as input for the embedding procedure and to the trigger selection of the $\tau$
decay products, now need to be applied.
The combined effect of these corrections is shown in Figure~\ref{fig:taucorr1} for the distributions of two quantities
closely related to their source: the pseudorapidity and the transverse momentum of the $\tau$ decay lepton.
The mismodelling of the pseudorapidity before corrections, cf.~Figure~\ref{fig:taucorr1a}, is due to detector acceptance
differences between the input muons and embedded $\tau$ objects.  While the corrections have a visible effect here, their
impact is found to be very small for the lepton $\pt$ shown in Figure~\ref{fig:taucorr1b} and also for any other of the
investigated quantities. Even after corrections, the modelling of the pseudorapidity is not perfect but, as demonstrated
below, this has no impact on observables relevant for physics analyses.

\begin{figure}[!h]
   \centering
            \subfloat[]{
        \includegraphics[width=0.50\textwidth]{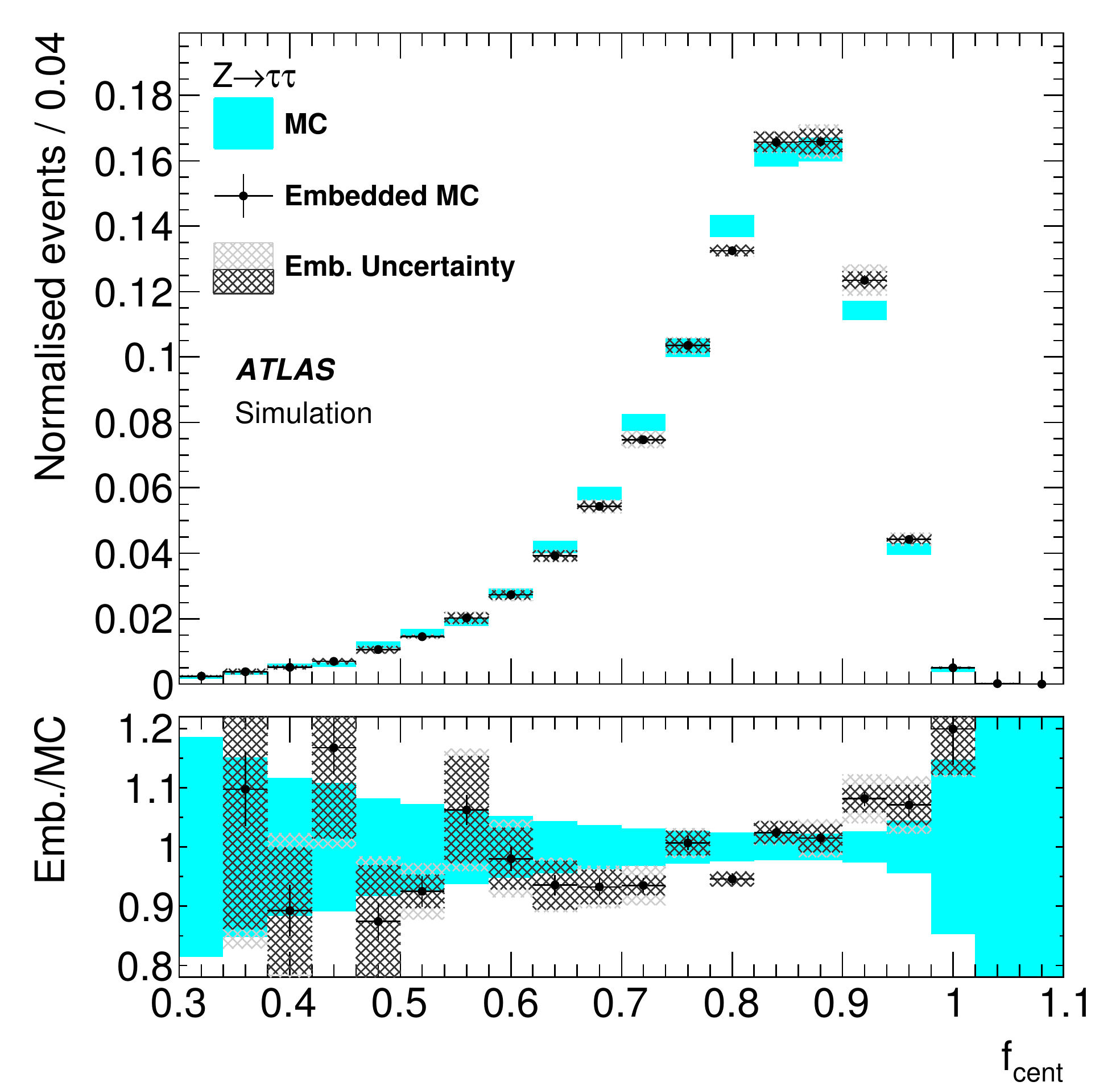}
        \label{fig:mcclosure1a}
   }
  \subfloat[]{
        \includegraphics[width=0.50\textwidth]{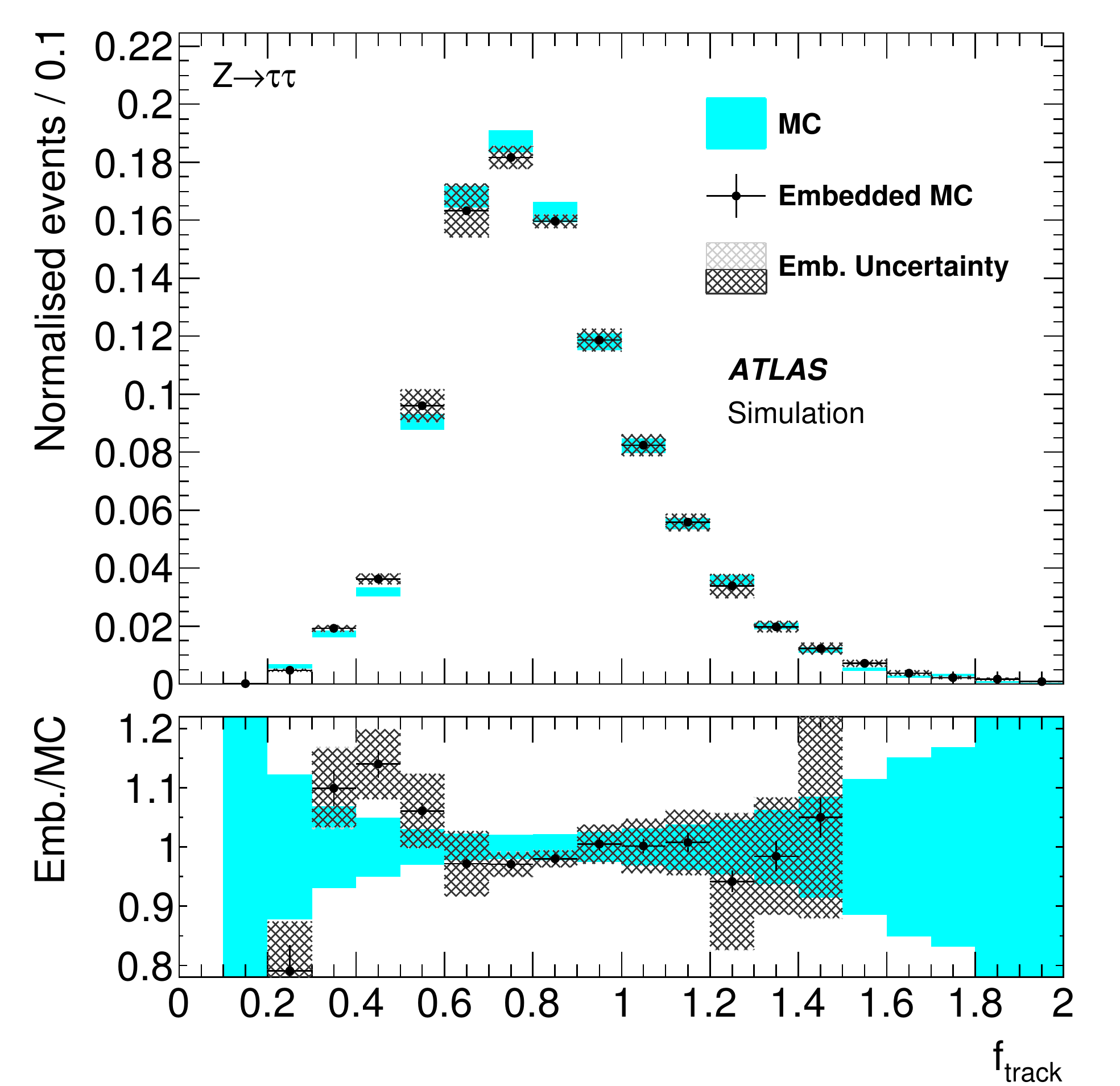}
        \label{fig:mcclosure1b}
   }\\
   \subfloat[]{
        \includegraphics[width=0.50\textwidth]{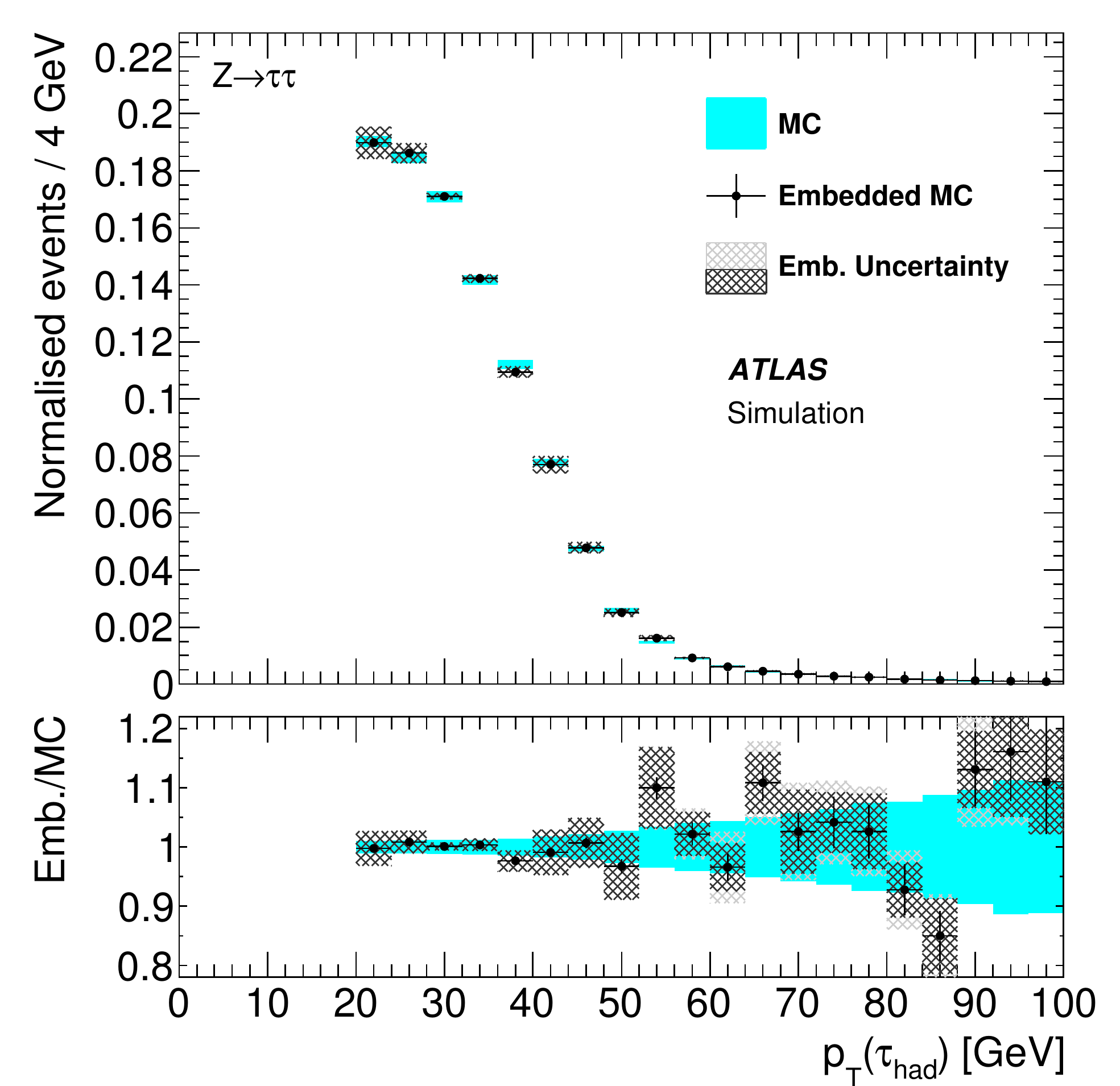}
        \label{fig:mcclosure1c}
   }
     \subfloat[]{
           \includegraphics[width=0.50\textwidth]{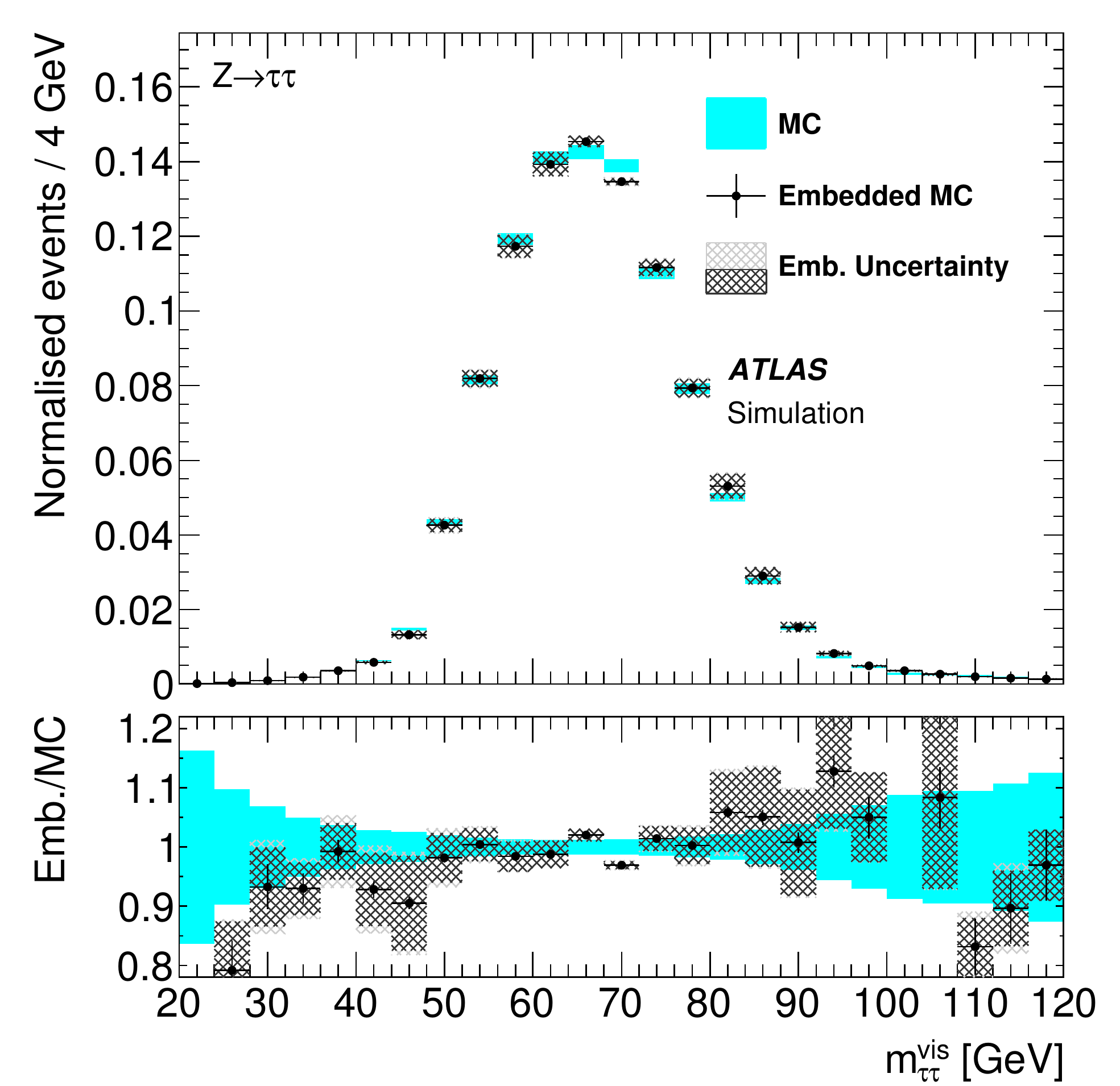}
           \label{fig:mcclosure1d}
   }
      \caption{
     \label{fig:mcclosure1}
             Comparison of $\tau$-embedded $Z\rightarrow\mu\mu$ MC events (black points) with $Z\rightarrow\tau\tau$
             MC events (blue):
              (a) central energy fraction,
              (b) leading-track momentum fraction for three-prong hadronic $\tau$ decays,
              (c) $\tau_\mathrm{had}$ transverse momentum and
              (d) mass of the visible $\tau\tau$ decay products, each including ratios showing the relative differences of the
              $\tau$-embedded distributions.
               The blue error band in the ratio plots corresponds to the statistical uncertainties of the $Z\rightarrow\tau\tau$ MC
               sample, and the black error bars are the statistical uncertainties associated with the $\tau$-embedded events. The
               light (dark) grey hatched error band corresponds to the sum in quadrature of cell + isolation (cell only) systematic
               uncertainties and the statistical uncertainties of the $\tau$-embedded events.
    }
\end{figure}

\begin{figure}[!h]
   \centering
            \subfloat[]{
        \includegraphics[width=0.50\textwidth]{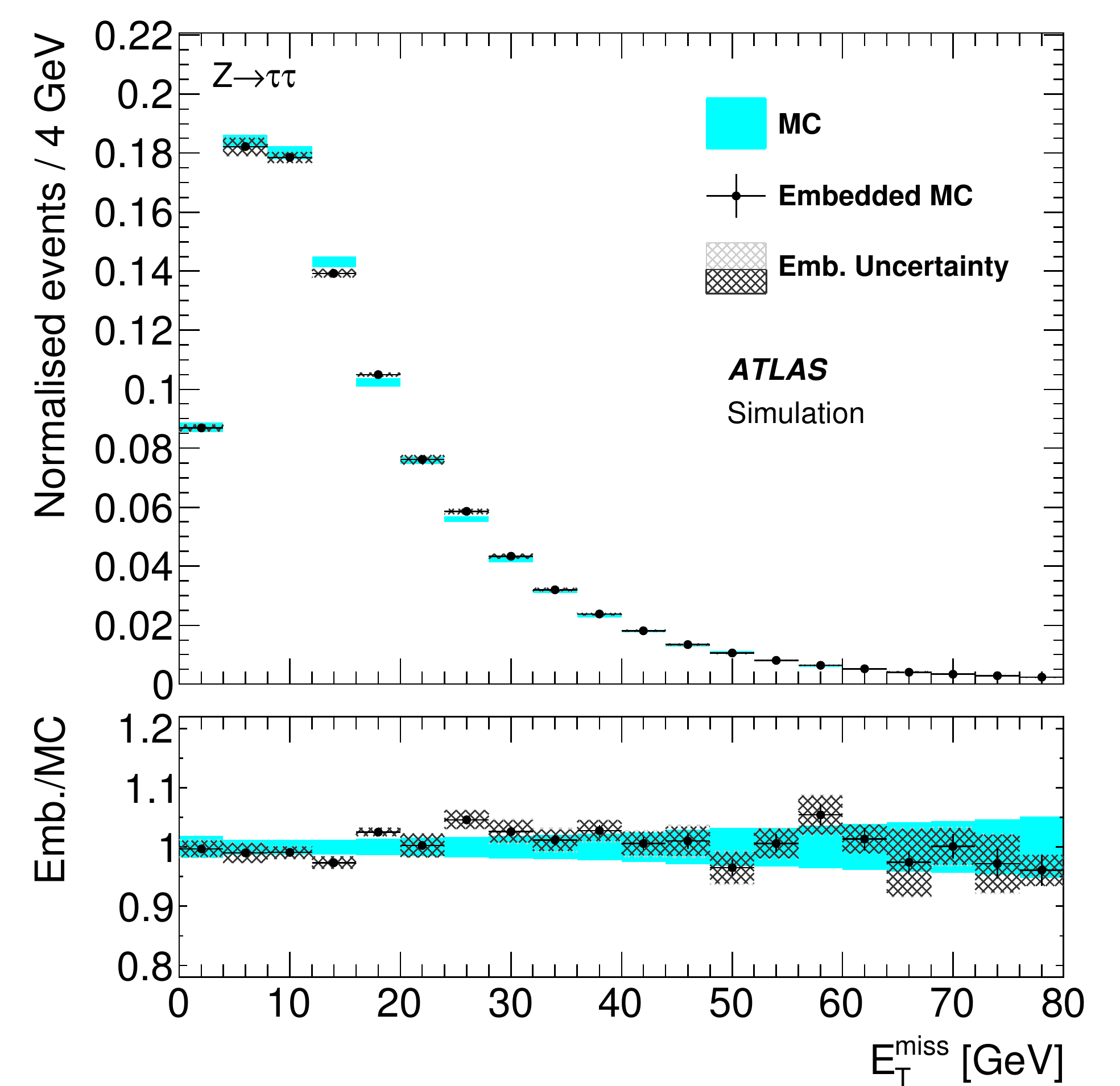}
        \label{fig:mcclosure2a}
   }
  \subfloat[]{
        \includegraphics[width=0.50\textwidth]{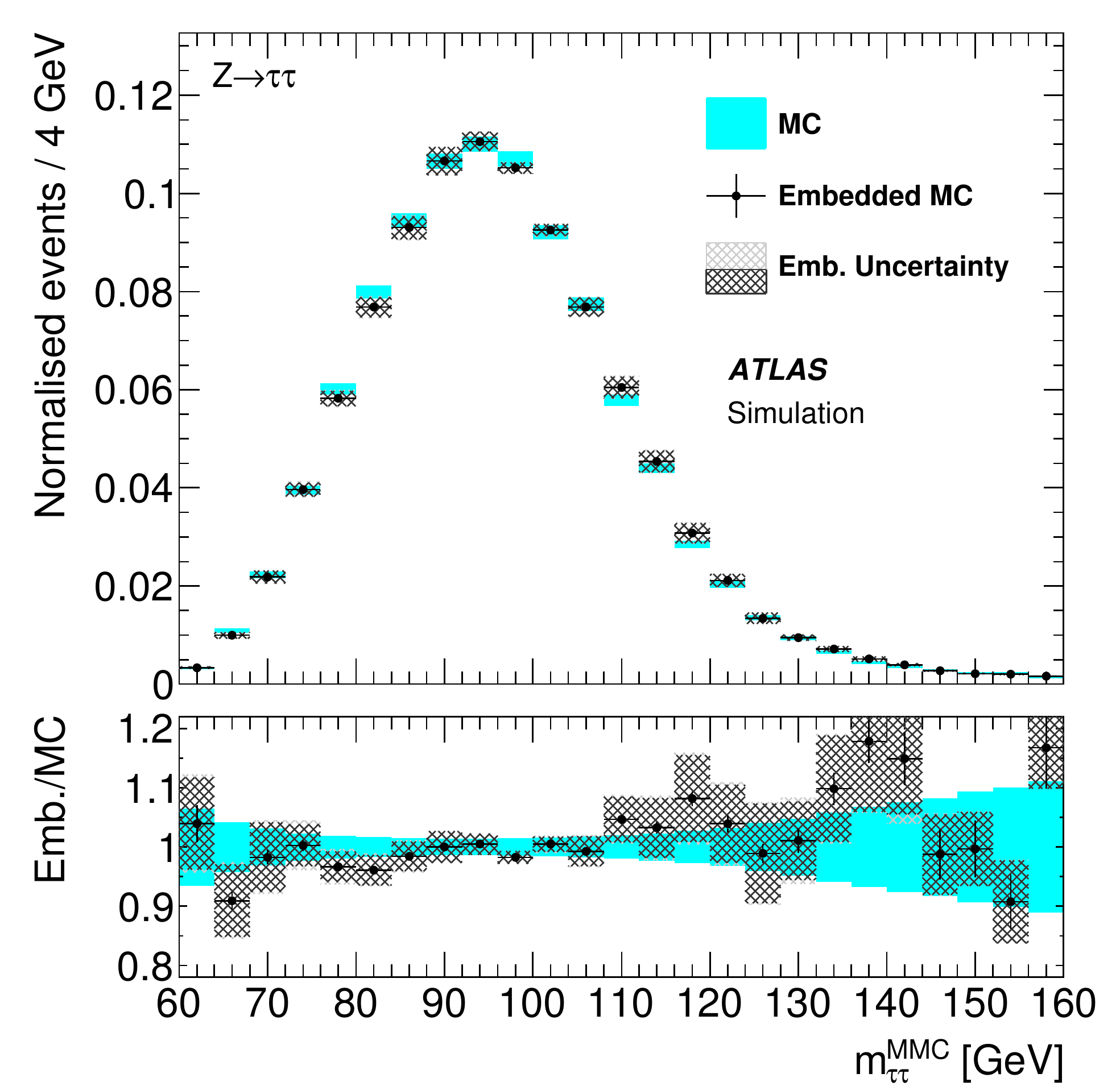}
     \label{fig:mcclosure2b}
   }\\
       \subfloat[]{
        \includegraphics[width=0.50\textwidth]{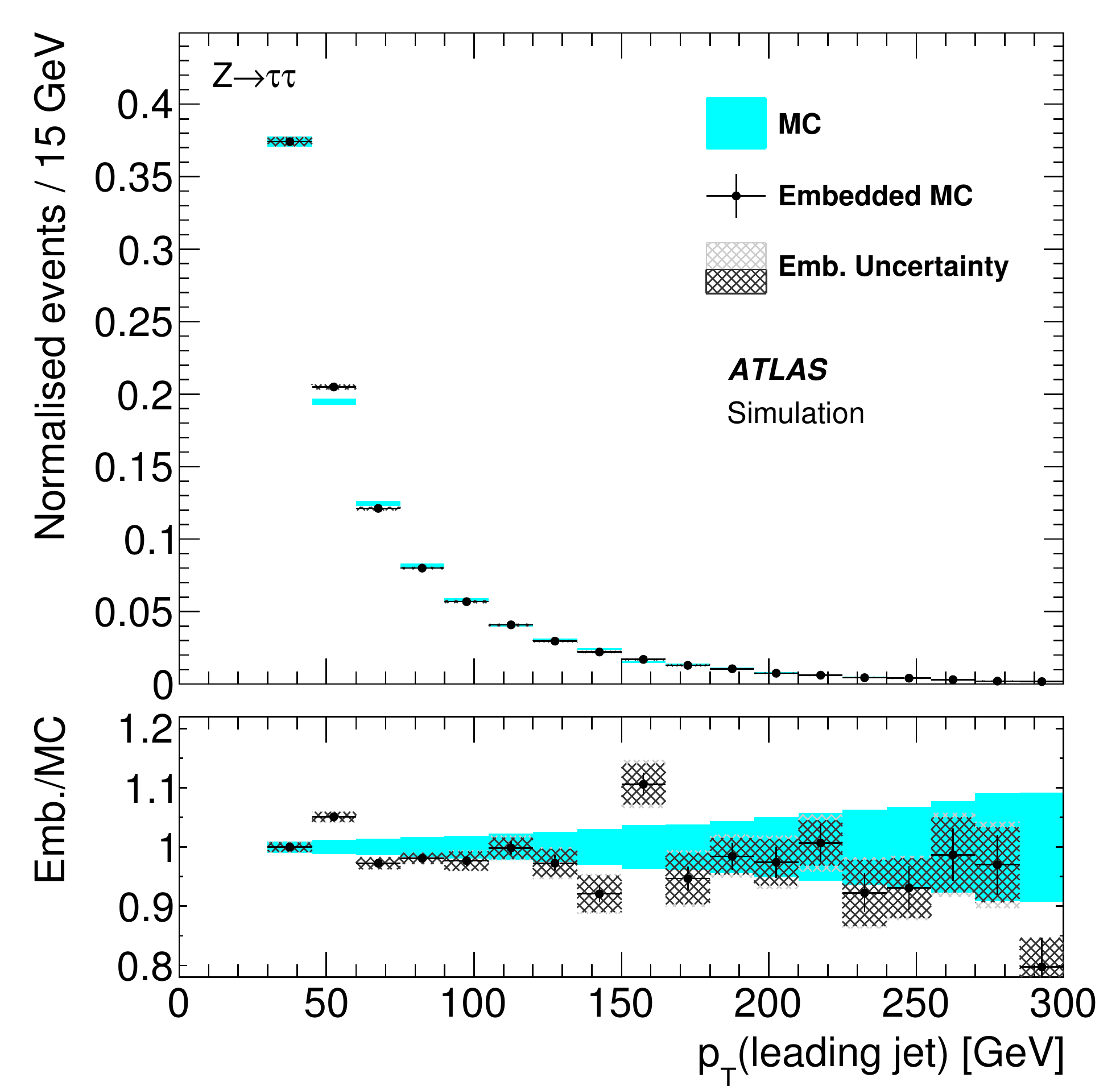}
         \label{fig:mcclosure2c}
   }
      \subfloat[]{
        \includegraphics[width=0.50\textwidth]{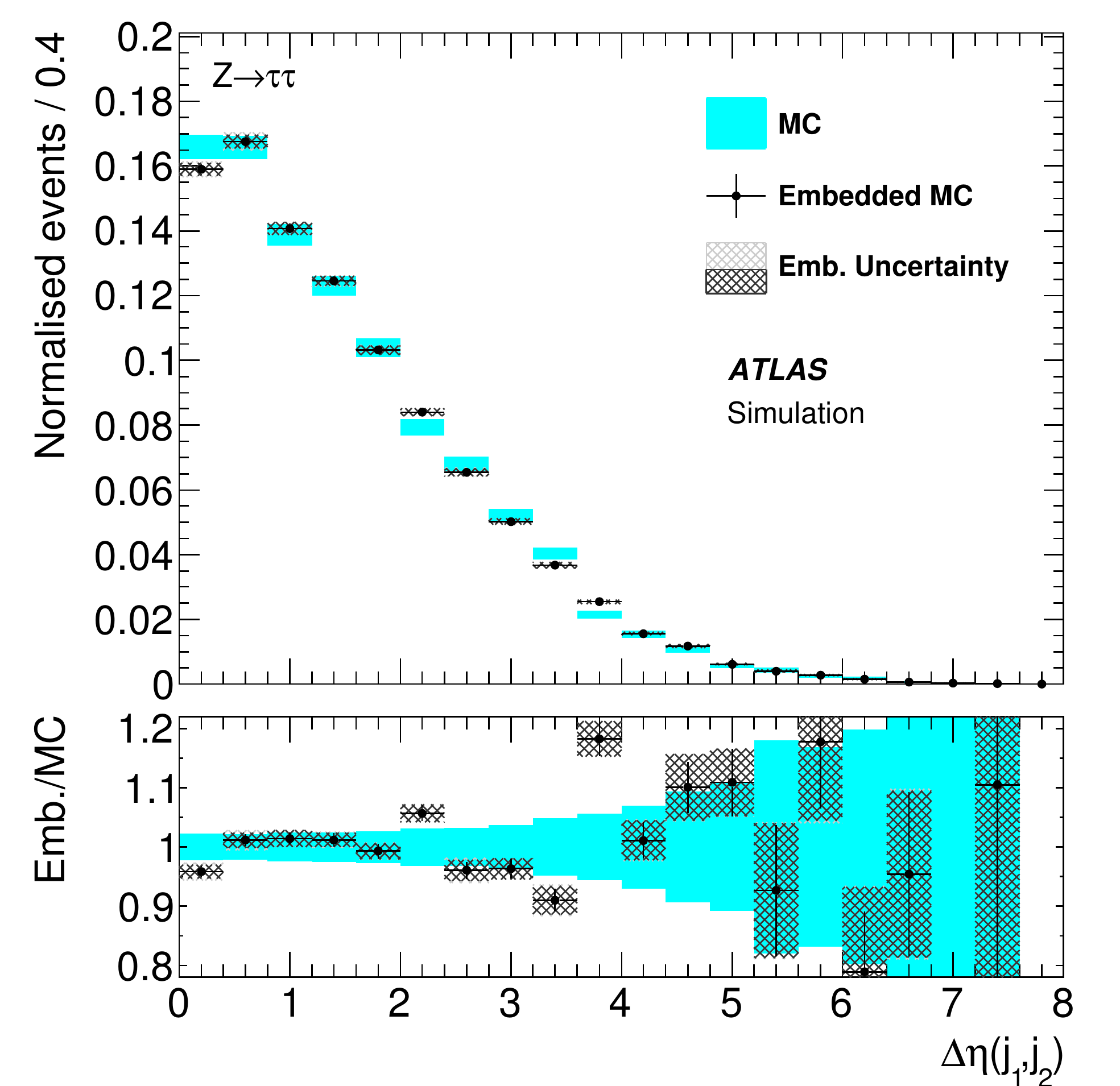}
         \label{fig:mcclosure2d}
   }
      \caption{
     \label{fig:mcclosure2}
             Comparison of $\tau$-embedded $Z\rightarrow\mu\mu$ MC events (black points) with $Z\rightarrow\tau\tau$
             MC events (blue):
                (a) missing transverse momentum,
                (b) $\tau\tau$ invariant mass obtained with the MMC,
                (c) transverse momentum of the leading jet and
               (d) pseudorapidity difference for the two leading jets, each including ratios showing the relative differences of the
               $\tau$-embedded distributions.
              The blue error band in the ratio plots corresponds to the statistical uncertainties of the $Z\rightarrow\tau\tau$ MC
              sample, and the black error bars are the statistical uncertainties associated with the $\tau$-embedded events. The
              light (dark) grey hatched error band corresponds to the sum in quadrature of cell + isolation (cell only) systematic
              uncertainties and the statistical uncertainties of the $\tau$-embedded events.
    }
\end{figure}

Further examples of such comparisons,
from here on omitting the uncorrected distributions, 
are collected in Figure~\ref{fig:mcclosure1} and Figure~\ref{fig:mcclosure2}.
Figures \ref{fig:mcclosure1a} and \ref{fig:mcclosure1b} show the distributions of two of the input quantities for the hadronic
$\tau$ identification: the central energy fraction, which is the ratio of the transverse energy deposited within $\Delta R$< 0.1
and $\Delta R$< 0.2 around the $\tau$ candidate direction, and the leading-track momentum fraction, i.e. the transverse
momentum of the highest-$\pt$ charged particle divided by the calorimetric transverse energy within
$\Delta R$< 0.2~\cite{Aad:2014rga}. Agreement of the distributions within statistical and embedding-related systematic
uncertainties indicates that the detector response to embedded $\tau$ leptons does not differ significantly from the standard
$Z\to\tau\tau$ MC samples. This is further confirmed by the fact that the $\tau$ identification efficiency is found to agree
for $\tau$-embedded and standard MC samples for all working points defined in Ref.~\cite{Aad:2014rga} within uncertainties.
Agreement is also observed for the kinematics of the $Z$ decay products, as demonstrated for the $\tau_{\rm had}$ \pt\ and
$m^{\mathrm{vis}}_{\tau\tau}$ in Figures \ref{fig:mcclosure1c} and  \ref{fig:mcclosure1d}.

Figures \ref{fig:mcclosure2a} and  \ref{fig:mcclosure2b}  compare the distributions of the missing transverse momentum,
arising from the simulated $\tau$ decay neutrinos and reconstruction effects, and of $m_{\tau\tau}^{\mathrm{MMC}}$.
Again, no significant differences are observed and the same conclusions are reached for jet-related quantities, such as
the leading-jet $\pt$ and the pseudorapidity separation of the two leading jets shown in Figures \ref{fig:mcclosure2c}
and \ref{fig:mcclosure2d} .

Thus, the $\tau$-embedded $Z\to\mu\mu$ MC events and standard $Z\to\tau\tau$ MC events are found to agree
in all distributions identified to be relevant for physics analyses within the statistical and embedding-related systematic
uncertainties described in Section~\ref{subsec:syst}.
These comparisons include effects from the modification of the input muon kinematics due to final-state radiation and
resolution and thus confirm that such effects are also covered by the current $\tau$-embedding uncertainties.

\clearpage

\subsection*{Performance within physics analyses}
\label{subsec:physval}
\begin{figure}[h!]
  \vspace*{0.25cm}
                \subfloat[]{
            \includegraphics[width=0.50\textwidth]{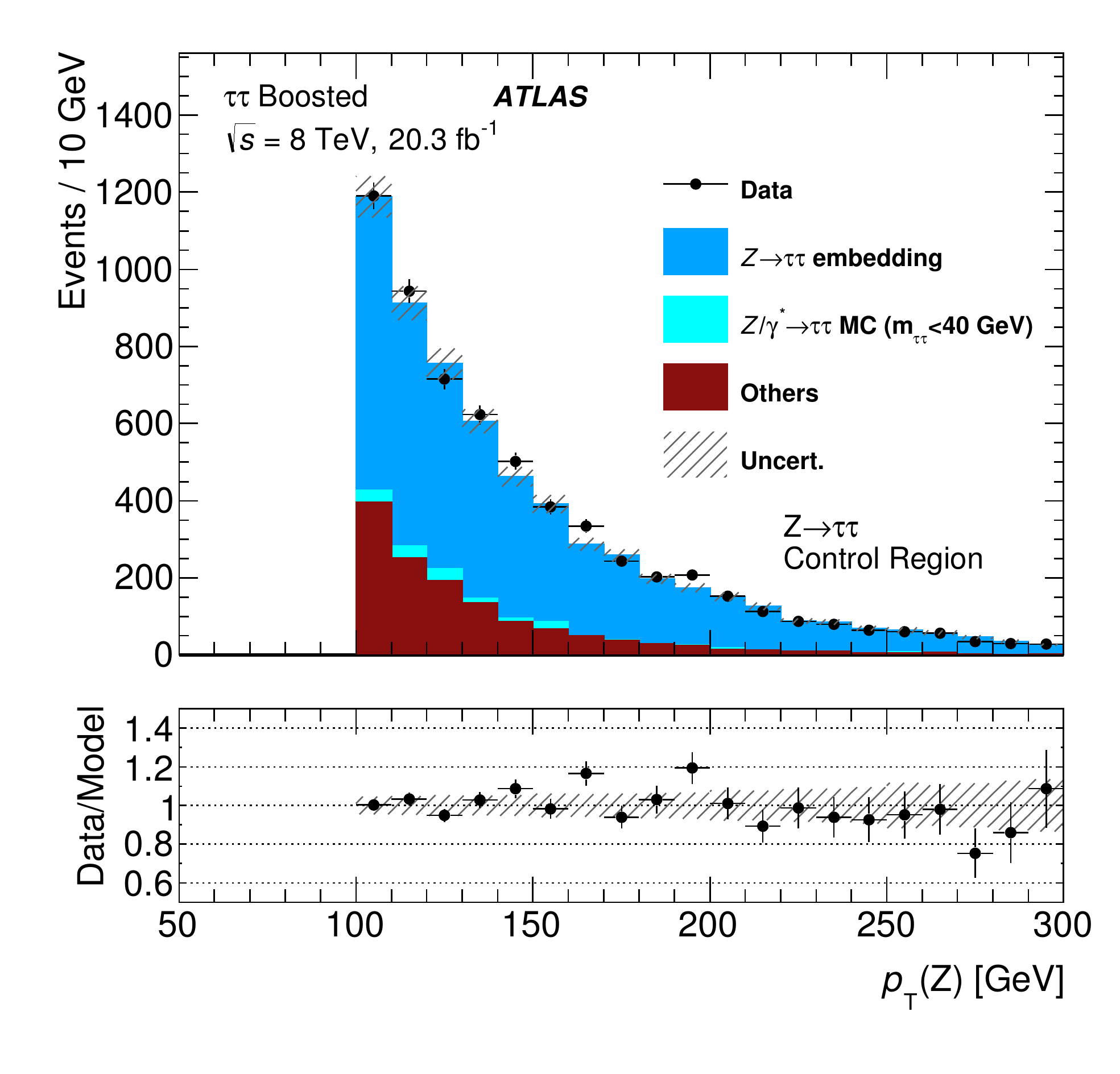}
             \label{fig:physval2a}
                     }
       \subfloat[]{
            \includegraphics[width=0.50\textwidth]{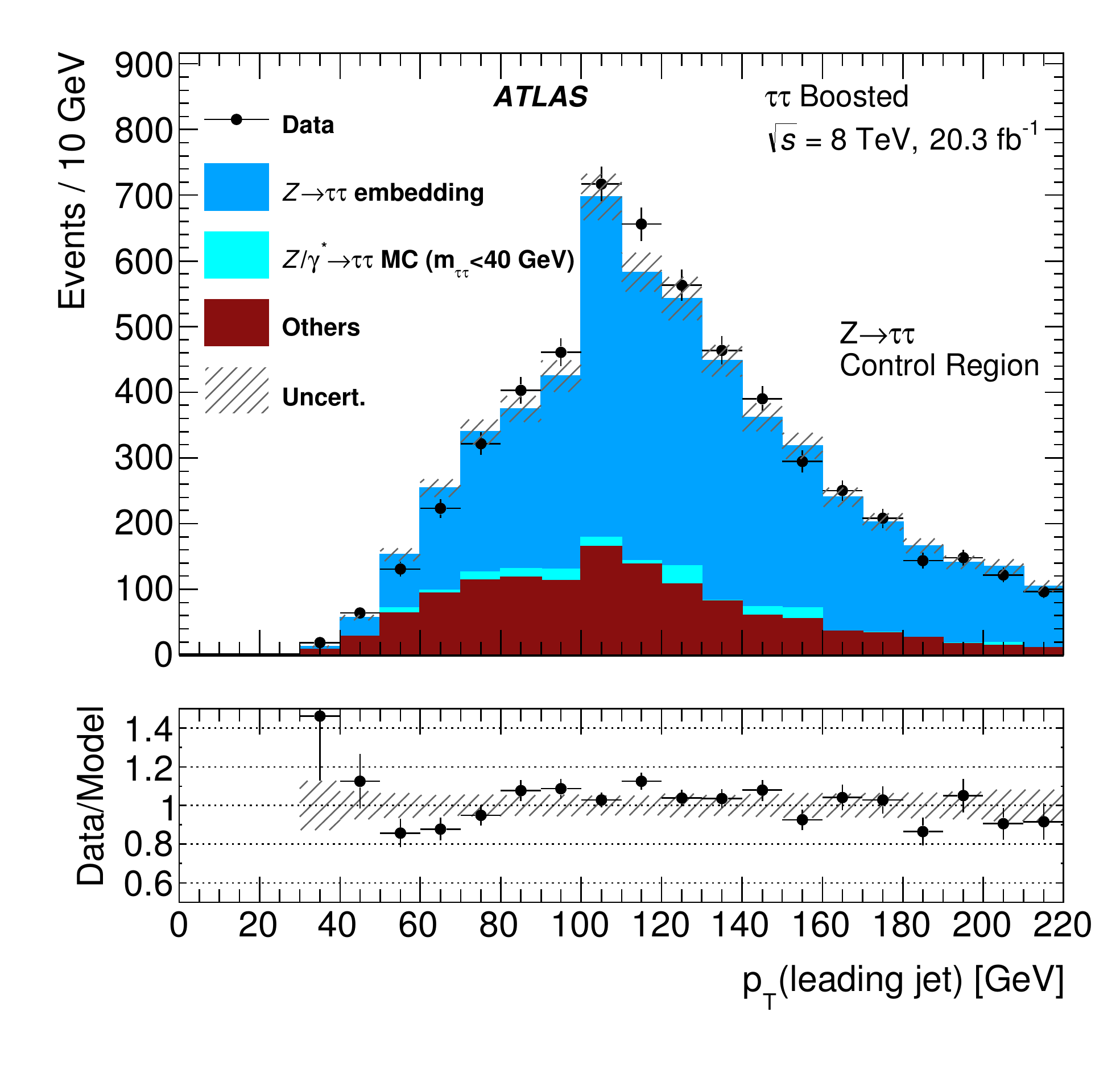}
             \label{fig:physval2b}
            }
             \caption{
            \label{fig:physval2}
            Comparison of data with the combined background model for example observables in the boosted $Z$-enriched
            control region:
transverse momentum of
 (a) the $Z$ boson  and (b)  the leading jet, each including ratios showing the relative differences of the data to the total
 background estimate. The background contributions from other processes and the systematic uncertainties are estimated
 as described in Ref.~\cite{Aad:2015vsa}.
            }
            \vspace*{0.5cm}
\end{figure}

%
In a final step, the $\tau$-embedded $Z\to\mu\mu$ collision data events are used as part of a combined background model
and compared to data in the boosted $Z$-enriched control region defined in Section~\ref{subsec:selection}.
\begin{figure}[b!]
     \vspace*{-0.2cm}
   \subfloat[]{
            \includegraphics[width=0.50\textwidth]{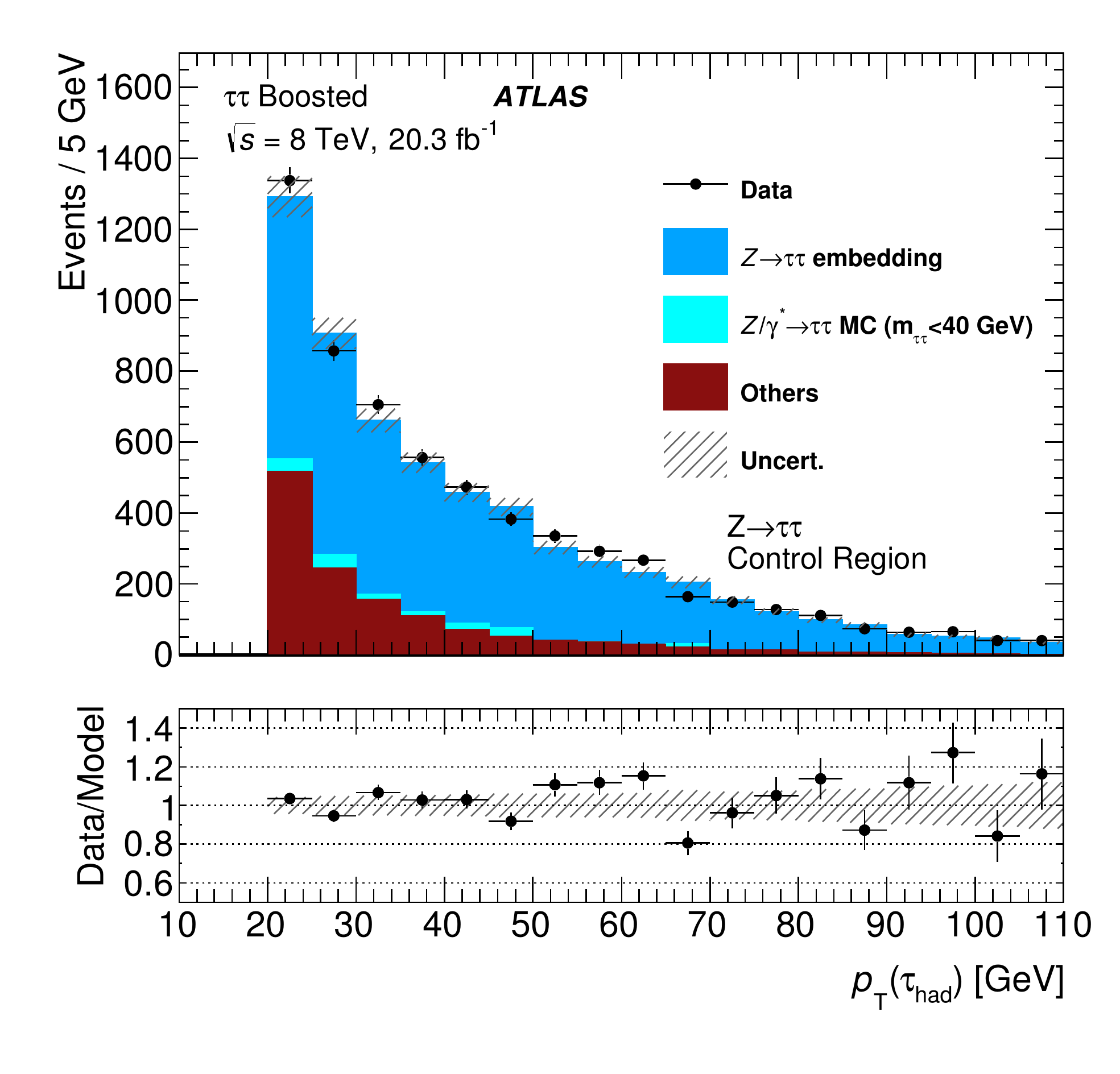}
            \label{fig:physval1a}
                    }
   \subfloat[]{
            \includegraphics[width=0.50\textwidth]{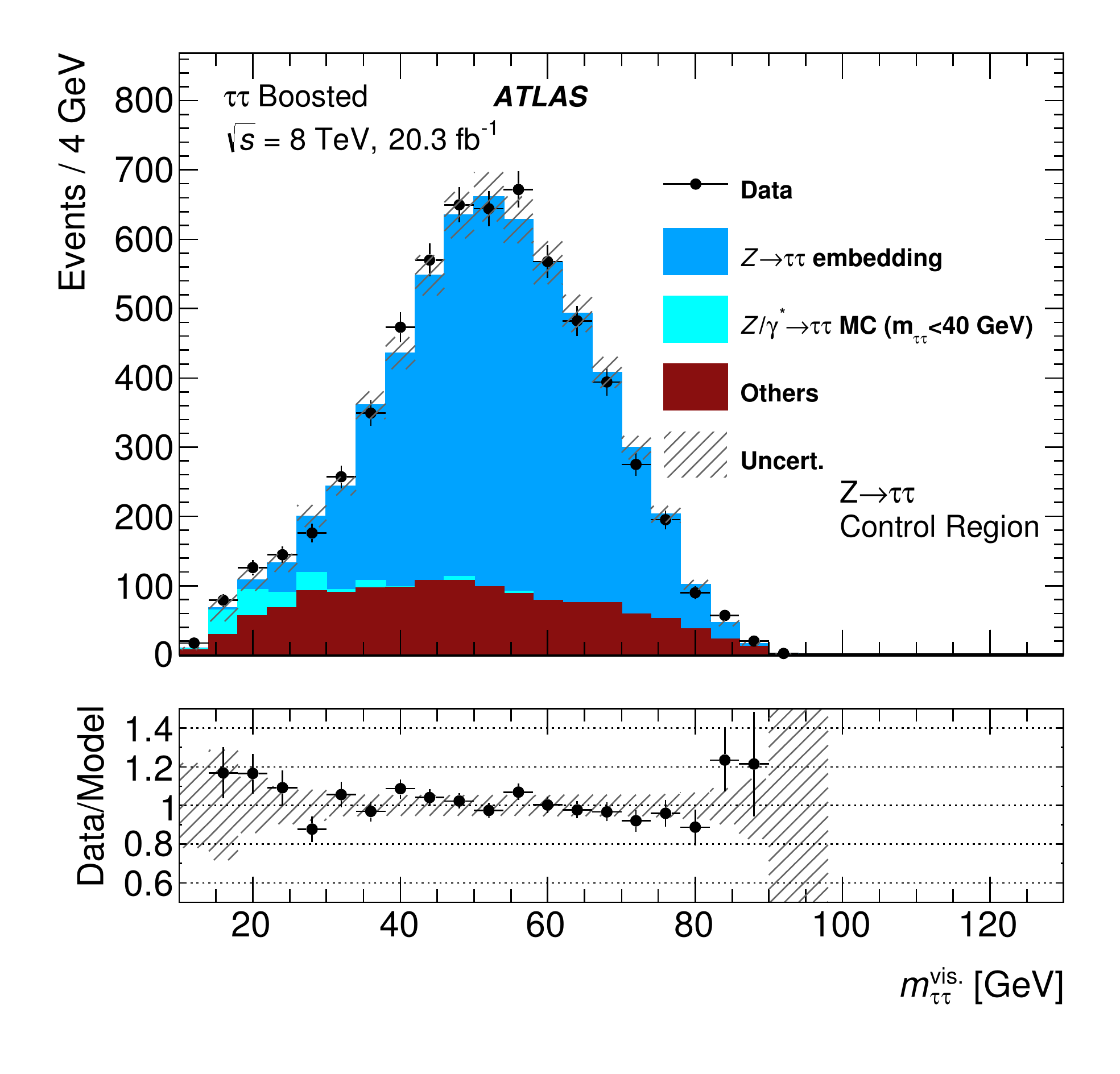}
            \label{fig:physval1b}
           }\\
                \subfloat[]{
            \includegraphics[width=0.50\textwidth]{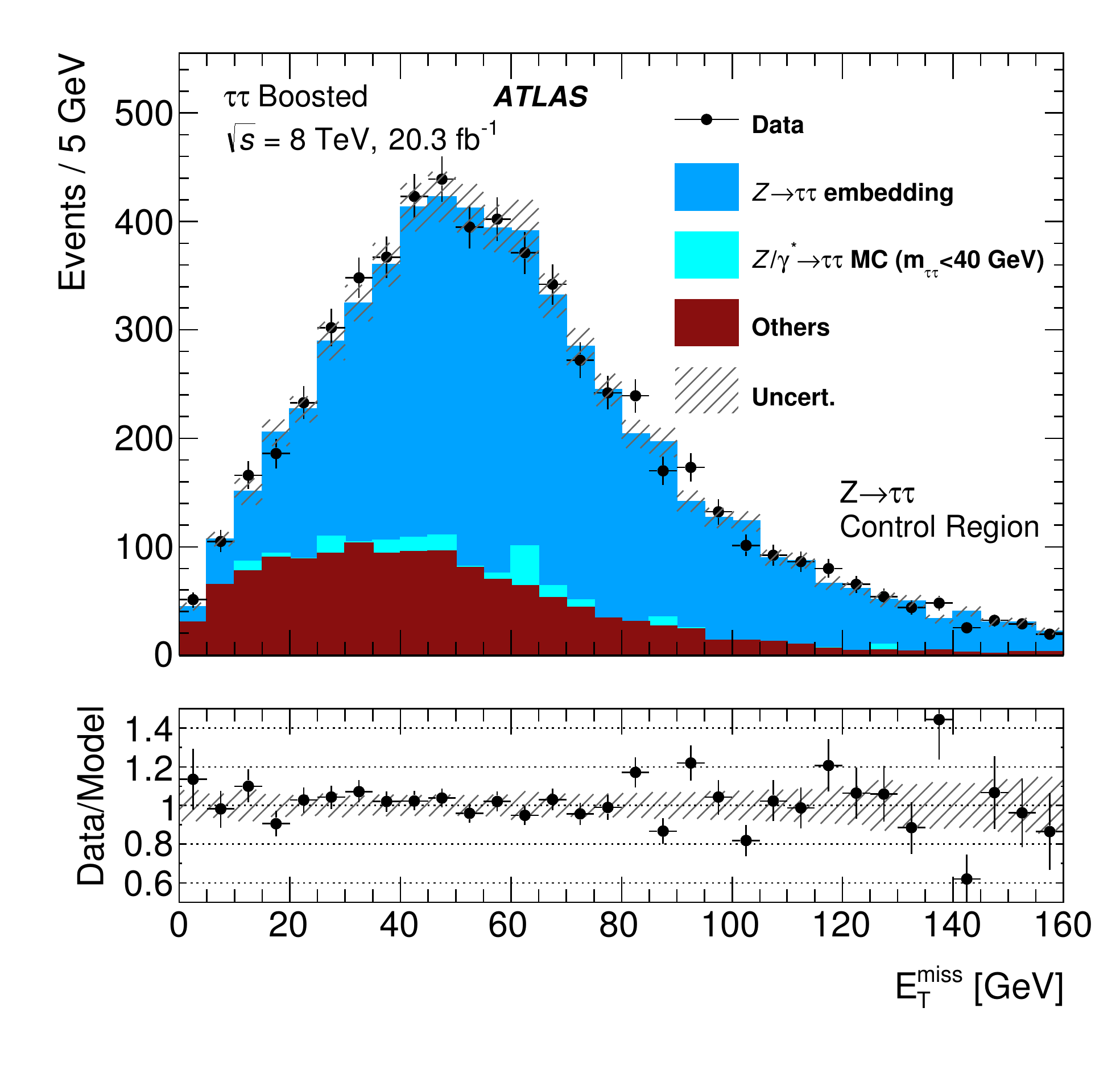}
            \label{fig:physval1c}
             }
                 \subfloat[]{
            \includegraphics[width=0.50\textwidth]{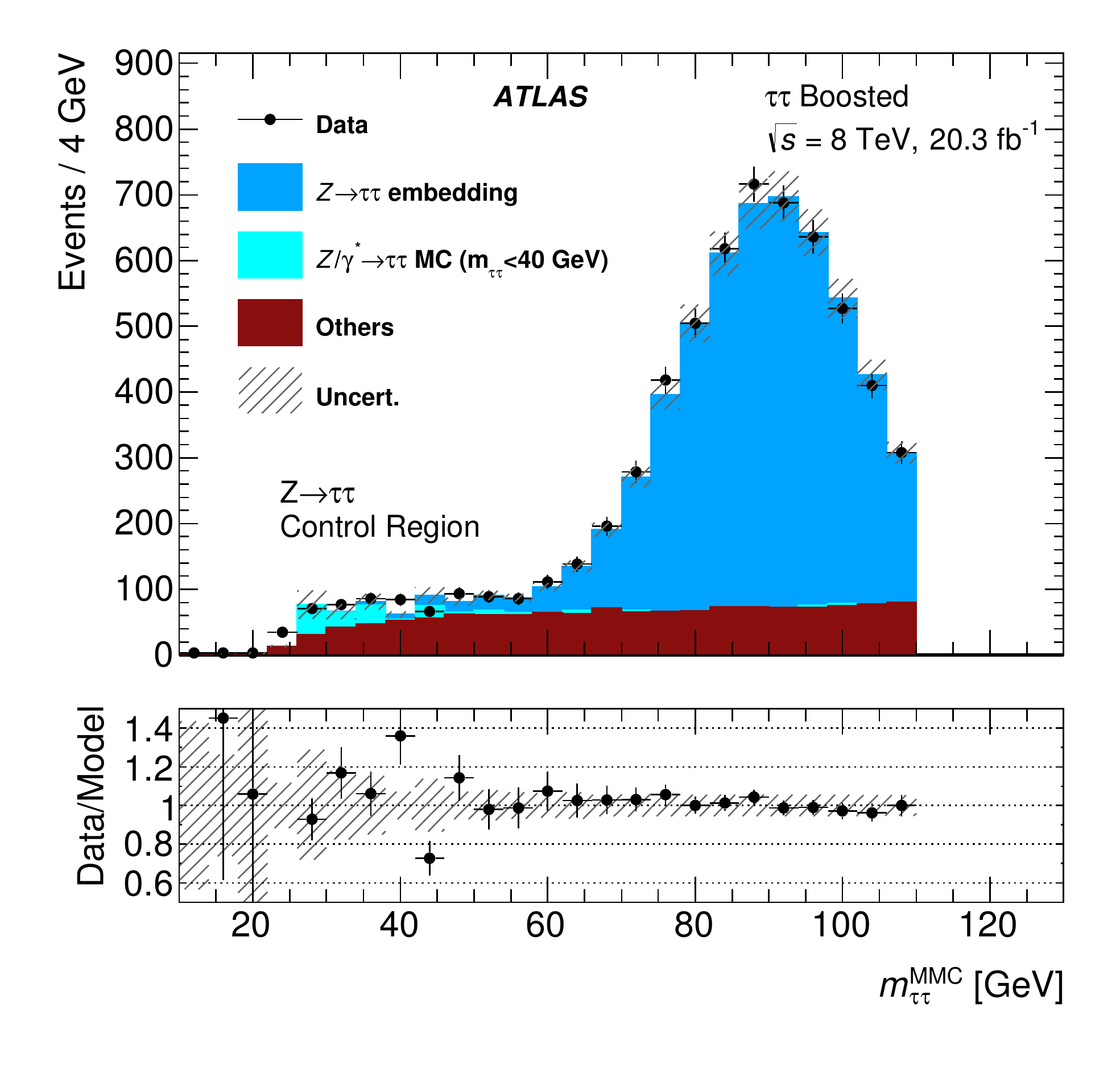}
            \label{fig:physval1d}
           }
              \caption{
            \label{fig:physval1}
            Comparison of data with the combined background model for example observables in the boosted $Z$-enriched
            control region:
           (a) $\tau_{\mathrm{had}}$ transverse momentum, (b) invariant mass of the visible $\tau\tau$ decay products
             (c) missing transverse momentum and
             (d) the $\tau\tau$ invariant mass obtained with the MMC, each including ratios showing the relative differences of
             the data to the total background estimate.
           The background contributions from other processes and the systematic uncertainties are estimated as described in
           Ref.~\cite{Aad:2015vsa}.
            }
\end{figure}
Due to significant contributions from other background processes, this is not a clean, stand-alone validation of the
embedding method but involves other background estimation procedures, performed exactly as in
Ref.~\cite{Aad:2015vsa}. Since the selection of  $Z\to\mu\mu$ data events used as input for the embedding procedure
includes a cut on the  invariant mass $m_{\mu\mu}>40$~GeV, low mass Drell--Yan processes with $\tau\tau$ final states
are not modelled via the embedding technique. Instead, these contributions are separately estimated from simulated
event samples.
A few example comparisons are given in Figures \ref{fig:physval1} and \ref{fig:physval2}.  In those distributions
the embedded samples are normalised to data in a dedicated region as described in Ref.~\cite{Aad:2015vsa}.
The combined background distributions, dominated by the embedding-based $Z\to\tau\tau$ model, are found to
provide a good description of the ATLAS data within the uncertainties, which here also include other relevant
uncertainties related to the estimation of the other background contributions as described in Ref.~\cite{Aad:2015vsa}.

\FloatBarrier
\section{Summary and conclusions}
\label{sec:conclusions}
%
This paper presented the motivation, concept and technical implementation of a $\tau$-embedding method, which models
events with $Z\to\tau\tau$ decays and possibly additional jets in a largely data-driven way. In $Z\to\mu\mu$ events
selected from $pp$ collision data recorded with the ATLAS experiment during the LHC Run1, tracks and calorimeter cell
energies associated with the $Z$ decay muons are replaced by the corresponding tracks and energy depositions of the
$\tau$ leptons from simulated $Z\to\tau\tau$ decays. For each event, the $\tau$ kinematics are derived from the original
$Z\to\mu\mu$ decay, so that their correlations with other event properties such as additional jets and the reconstructed
missing transverse momentum are preserved in the resulting hybrid $Z\to\tau\tau$ events. Systematic uncertainties are
estimated by varying the muon isolation requirement and the subtracted energy depositions associated with the muons.
Extensive validation studies were performed using both the $\mu\mu$ and $\tau\tau$ final states, presented here only for
the example where one $\tau$ lepton decays into an electron or muon and the other hadronically. The $\mu\mu$-based
results demonstrate that the procedure successfully replaces objects in the data events without affecting other event
properties. Comparing $\tau$-embedded $Z\to\mu\mu$ MC events with standard $Z\to\tau\tau$ MC, agreement was
found for distributions of all quantities relevant to current physics analyses within the combined statistical and
embedding-related systematic uncertainties. Other conceptual limitations of the method related to the input of reconstructed
muon kinematics are found to introduce only small effects compared to the uncertainties estimated from variations of
the method. For Higgs analyses in $\tau\tau$ final states, which exploit intricate signatures of additional jets and their
correlation with the $\tau\tau$ decay kinematics, the $\tau$-embedded data thus provide a reliable model of the irreducible
background from events with $Z\to\tau\tau$ decays and jets.
\vspace*{0.5cm}
\section*{Acknowledgements}
\label{sec:acknowldge}
%
We thank CERN for the very successful operation of the LHC, as well as the
support staff from our institutions without whom ATLAS could not be
operated efficiently.

We acknowledge the support of ANPCyT, Argentina; YerPhI, Armenia; ARC,
Australia; BMWFW and FWF, Austria; ANAS, Azerbaijan; SSTC, Belarus; CNPq and FAPESP,
Brazil; NSERC, NRC and CFI, Canada; CERN; CONICYT, Chile; CAS, MOST and NSFC,
China; COLCIENCIAS, Colombia; MSMT CR, MPO CR and VSC CR, Czech Republic;
DNRF, DNSRC and Lundbeck Foundation, Denmark; EPLANET, ERC and NSRF, European Union;
IN2P3-CNRS, CEA-DSM/IRFU, France; GNSF, Georgia; BMBF, DFG, HGF, MPG and AvH
Foundation, Germany; GSRT and NSRF, Greece; RGC, Hong Kong SAR, China; ISF, MINERVA, GIF, I-CORE and Benoziyo Center, Israel; INFN, Italy; MEXT and JSPS, Japan; CNRST, Morocco; FOM and NWO, Netherlands; BRF and RCN, Norway; MNiSW and NCN, Poland; GRICES and FCT, Portugal; MNE/IFA, Romania; MES of Russia and NRC KI, Russian Federation; JINR; MSTD,
Serbia; MSSR, Slovakia; ARRS and MIZ\v{S}, Slovenia; DST/NRF, South Africa;
MINECO, Spain; SRC and Wallenberg Foundation, Sweden; SER, SNSF and Cantons of
Bern and Geneva, Switzerland; NSC, Taiwan; TAEK, Turkey; STFC, the Royal
Society and Leverhulme Trust, United Kingdom; DOE and NSF, United States of
America.

The crucial computing support from all WLCG partners is acknowledged
gratefully, in particular from CERN and the ATLAS Tier-1 facilities at
TRIUMF (Canada), NDGF (Denmark, Norway, Sweden), CC-IN2P3 (France),
KIT/GridKA (Germany), INFN-CNAF (Italy), NL-T1 (Netherlands), PIC (Spain),
ASGC (Taiwan), RAL (UK) and BNL (USA) and in the Tier-2 facilities
worldwide.
%
\newpage
\printbibliography
%
\newpage
\begin{flushleft}
{\Large The ATLAS Collaboration}

\bigskip

G.~Aad$^{\rm 85}$,
B.~Abbott$^{\rm 113}$,
J.~Abdallah$^{\rm 151}$,
O.~Abdinov$^{\rm 11}$,
R.~Aben$^{\rm 107}$,
M.~Abolins$^{\rm 90}$,
O.S.~AbouZeid$^{\rm 158}$,
H.~Abramowicz$^{\rm 153}$,
H.~Abreu$^{\rm 152}$,
R.~Abreu$^{\rm 116}$
% ...
\bigskip
\\
% ...
$^{11}$ Institute of Physics, Azerbaijan Academy of Sciences, Baku, Azerbaijan\\
% ...
$^{85}$ CPPM, Aix-Marseille Universit{\'e} and CNRS/IN2P3, Marseille, France\\
% ...
$^{90}$ Department of Physics and Astronomy, Michigan State University, East Lansing MI, United States of America\\
% ...
$^{107}$ Nikhef National Institute for Subatomic Physics and University of Amsterdam, Amsterdam, Netherlands\\
% ...
$^{113}$ Homer L. Dodge Department of Physics and Astronomy, University of Oklahoma, Norman OK, United States of America\\
% ...
$^{116}$ Center for High Energy Physics, University of Oregon, Eugene OR, United States of America\\
% ...
$^{151}$ Institute of Physics, Academia Sinica, Taipei, Taiwan\\
$^{152}$ Department of Physics, Technion: Israel Institute of Technology, Haifa, Israel\\
$^{153}$ Raymond and Beverly Sackler School of Physics and Astronomy, Tel Aviv University, Tel Aviv, Israel\\
% ...
$^{158}$ Department of Physics, University of Toronto, Toronto ON, Canada\\
% ...
\end{flushleft}


\end{document}